\documentclass[final,5p,times,twocolumn]{elsarticle}

\usepackage{newsuse}
\usepackage[USenglish]{babel}
\usepackage[final]{pdfpages}
\graphicspath{{./figures/}}

\usepackage{xurl}
\usepackage{xcolor}

\begin{document}
\begin{frontmatter}



\title{\LARGE AI-Powered Browsers Are Broadly Accurate News Summarizers That Reduce Political Bias and Negative Affect}


\author[first]{Yan Xia}
\author[second]{Dominik Batorski}
\author[first]{Erin Wertz}
\author[first]{Michalis Mamakos}
\author[third]{Lucen Li}
\author[first,third]{Magdalena Wojcieszak\corref{cor1}}

\cortext[cor1]{%
    Corresponding author. Email: 
    \href{mailto:mwojcieszak@ucdavis.edu}{mwojcieszak@ucdavis.edu}
    }

\affiliation[first]{organization={Center for Excellence in Social Science, University of Warsaw}, 
            country={Poland}}
\affiliation[second]{organization={Interdisciplinary Centre for Mathematical and Computational Modelling, University of Warsaw}, 
            country={Poland}}
\affiliation[third]{organization={Department of Communication, University of California, Davis}, 
            country={United States}}
    \begin{abstract}
Web browsers now provide AI-generated news summaries for millions of users. Despite their popularity and influence, we lack a systematic understanding of how these systems transform news before people read it. Through a large-scale audit, we investigate the factual accuracy of browser-based AI summarizers and how they alter the political bias, negative affect, and journalistic writing quality of news. Drawing on 13,777 articles from 15 U.S. news outlets, we evaluate their 41,331 summaries generated by three leading AI-powered browsers: Google Chrome (Gemini), Microsoft Edge (Copilot), and Perplexity Comet. We find that browser-based AI summarizers are broadly accurate. Furthermore, they consistently transform news by attenuating ideological bias, partisan stances, negativity, anger, and fear, while increasing clarity and reducing personal tone. With some variations, these patterns hold across browsers, outlet ideologies, and topics. Our findings identify AI-powered browsers as a new class of editorial intermediaries that systematically reshape news, with implications for democratic discourse and AI governance.

\end{abstract}

\end{frontmatter}



Information intermediaries play a key role in shaping how citizens learn about public affairs. Extensive research documents how search engines, news aggregators, social media platforms, and algorithmic recommendation systems influence the content users encounter online \cite{oecd2026ai,microsoft2026global,pew2025us,eurostat2025eu}. Yet, a new generation of AI-powered intermediaries is now emerging. Large language models (LLMs) are increasingly embedded into everyday information environments, including search, messaging applications, productivity software, and web browsers \cite{google2025go, microsoft2025meet}. Unlike the earlier intermediaries, which merely influence \textit{which} information users see, LLMs additionally  determine \textit{how} that information is presented, altering it before exposure occurs. By generating summaries, answers, explanations, and synthetic overviews, AI systems rewrite, condense, and reinterpret source material, creating a new layer of mediation through transformation, not merely exposure. 

This shift is particularly consequential in the domain of news and politics. News use through AI interfaces is rapidly increasing, especially among younger users, and AI summarization in particular is seen as the most appealing AI feature to news consumers \cite{newman2025reuters}.
Furthermore, AI summarization is becoming directly integrated into web browsers, the primary gateway to news online \cite{wojcieszak2022avenues, stier2022post}. Mainstream browsers such as Google Chrome and Microsoft Edge, as well as AI-native browsers such as Perplexity Comet, now allow users to generate summaries of news articles without leaving the webpage, potentially reshaping news consumption for millions of readers. 

Although these tools can make news more accessible and easier to digest, it is unclear whether they produce reliable news summaries. This question is crucial given the well-documented biases in LLMs \cite{gallegos2024bias,navigli2023biases, hofmann2024ai,abid2021persistent, kotek2023gender, rozado2024political,buyl2026large}, their failures in generic summarization tasks \cite{maynez2020faithfulness,ji2023survey}, and---concurrently---their power to influence users’ political attitudes \cite{fisher2025biased, bai2025explicitly, potter2024hidden, linPersuadingVotersUsing2025}. A news summary that introduces unsupported claims, amplifies anger, or increases ideological bias may breed misperceptions or deepen partisan divides. These risks are amplified by how summarization tools operate: these proprietary, black-box systems are designed to eliminate the need to read full articles, leaving users little reason or means to identify inaccuracies \cite{pew2025google}. 

Yet despite the growing importance of AI news summarization and its potential consequences, little systematic evidence exists on how these systems transform news content in practice. Existing audits primarily focus on accuracy, suggesting that AI chatbots can introduce factual errors, distort source material, or fail to attribute information correctly when asked questions about the news \cite{bbc2025ai, ebu2025news}. Although these findings raise important concerns, it is unclear whether and how AI summarization tools reshape the political, emotional, and stylistic characteristics of the news that users encounter. Key questions remain unaddressed: Do AI summarization tools amplify negativity, outgroup hatred, or partisan bias? Do they generate clearer or more objective summaries relative to the source articles? Are these tools better at summarizing partisan articles than centrist ones? 

We present the first large-scale audit of AI news summarization features deployed in real browsers, systematically evaluating not only their factual accuracy, but also how they alter the political bias, negative affect, and journalistic writing quality. We audit two mainstream browsers with built-in AI summarization, Google Chrome (powered by Gemini) and Microsoft Edge (powered by Copilot), alongside Perplexity Comet, the leading AI-native browser \cite{presenc2026perplexity}. We sampled 15,000 news articles from 15 U.S. news outlets that span the political spectrum (5 left-leaning, 5 centrist, and 5 right-leaning, 1,000 articles per outlet), and successfully retrieved AI-generated summaries from all three browsers for 13,777 articles, resulting in a total of 41,331 summaries; \ref{app:sec:data} provides details on outlet selection and data collection. Using these data, we first assess the factual accuracy of summaries against their source articles and analyze why inaccuracies arise. Second, we examine how AI summarization shifts ideological bias, partisan stance, negativity, anger, fear, as well as writing clarity, engagement, personal tone, and clickbait tendency (see~\ref{app:sec:data-annt} for details on measurement and validation). Third, we investigate how these patterns vary by the particular browser used, the leaning of the source outlet, as well as the topic of the article, ranging from the U.S. election to international conflicts (see~\ref{app:sec:topic} for details on the topic model and the 19 topics analyzed).

In so doing, we offer previously unavailable evidence in terms of both depth and scale. First, we move beyond accuracy and examine multiple dimensions of news content simultaneously. The political, emotional, and stylistic outcomes we examine
not only comprehensively capture how AI systems reshape news, but also bear directly on longstanding concerns about polarization \cite{mason2022uncivil, kalmoe2022radical}, hyper-partisan media \cite{yu2026us, levendusky2013partisan}, as well as news avoidance and political disengagement \cite{wertz2026correlates, villi2022taking}. 
Second, we advance research on algorithmic fairness by testing whether AI-driven news transformations are symmetric across the ideological spectrum. There is mounting evidence of political bias in algorithmic systems \cite{ibrahim_youtubes_2023, haroon2023auditing, lutz_examining_2021, bar2024systematic, huszar_algorithmic_2022, ye_auditing_2025} and LLM chatbots \cite{westwood_measuring_2025, potter2024hidden, bang_measuring_2024,buyl2026large,rozado2024political}; yet apart from a recent study suggesting that summaries by AI chatbots do differ for news originating from liberal versus conservative outlets \cite{savgira2026stays}, political bias in AI news summarization tools specifically remains largely underexplored. 

Third, we study AI summarization as users actually experience it. Past work evaluates model capabilities in controlled settings, by probing standalone chatbots with either news questions \cite{bbc2025ai,ebu2025news,suzgun2026evaluating} or custom summarization prompts \cite{alessa2025quantifying,savgira2026stays}. Such approaches help us understand model behavior but cannot capture how AI mediates news consumption in its arguably more consequential form: a dedicated, user-facing summarization feature in browsers, where users can ask the embedded AI to summarize a news webpage as they browse. By auditing three browser-integrated AI summarizers, we provide a more ecologically valid assessment of the information intermediaries citizens increasingly encounter in daily news use.

Despite widespread scholarly and public concern about AI tools, we find that browser-integrated AI news summarizers 
perform much better than expected, though not without caveats.
We offer four core findings. 
First, an average AI-generated summary of a news article is 82.8\% accurate, with errors mostly stemming from imprecise wording or supplementary context (see ~\ref{app:sec:res-acc} for our qualitative analysis of these errors). Second, AI summarizers consistently reduce political bias and negative affect in news articles. On average, 30.8\% of articles with ideological bias, 33.0\% with a partisan stance, 34.8\% with negativity, 56.1\% with anger, and 21.2\% with fear receive neutral summaries, whereas neutral articles much less often receive summaries with introduced ideological (2.6\%) or partisan bias (1.6\%), negativity (8.8\%), anger (2.3\%), or fear (11.7\%). 
However, the reduction in partisan bias is not uniform across leanings and tools: pro-Republican articles are significantly more likely to receive more neutral summaries than anti-Republican articles, and Edge 
attenuates right-leaning and pro-Republican stances more aggressively than other browsers (see ~\ref{app:sec:res-ideo} for our qualitative analysis of these patterns). 
Third, AI-generated news summaries exhibit higher journalistic writing quality than their source articles in terms of increased clarity, tempered personal tone, and fewer instances of clickbait, although engagement flattens under the summary format. 
Lastly, we show that these patterns---accurate summarization, reduced partisan bias, reduced negative affect, and largely improved journalistic writing quality---are robust across browsers, outlet leanings, and topics, despite variation in magnitude.

Taken together, our large-scale audit shows how AI-summarization tools transform political information across 11 outcomes, 15 news outlets, 19 salient topics, and three real-world deployment environments. By testing whether and how AI systems reshape the content users receive across browsers, outlet ideology, or political issues, we expand the growing literature on algorithmic mediation.

\section*{Results}\label{{sec:results}}

We analyze 13,777 news articles and their 41,331 summaries generated in Google Chrome, Microsoft Edge, and Perplexity Comet. The Materials and Methods section below provides the most important information on data collection, annotation, and validation; the appendices offer more in-depth details.

\begin{figure*}[htbp]
\centering
\includegraphics[width=\linewidth]{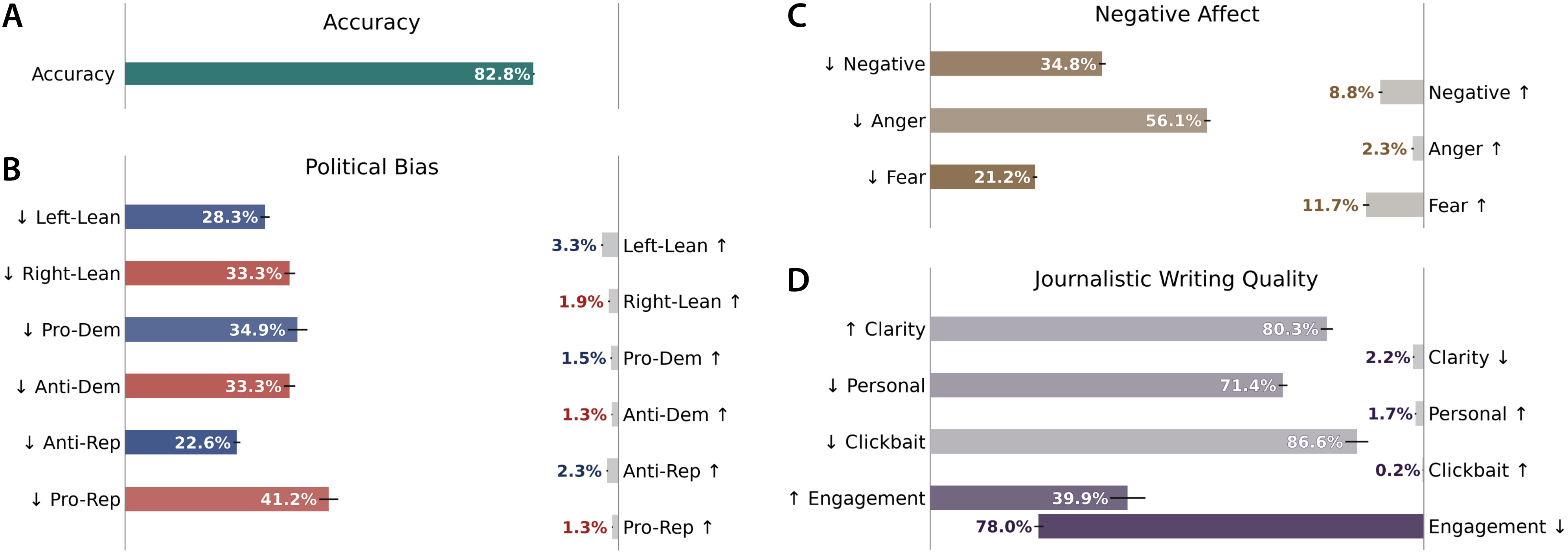}
\caption{Performance of AI-generated news summaries in terms of (A) factual accuracy, (B) change in political bias (ideological bias, and stance toward Democrats and Republicans), (C) change in negative affect (negativity, anger, and fear), and (D) change in journalistic writing quality (clarity, personal tone, clickbait, and engagement). In (B–D), bars extending from the left show the share of summaries in which each attribute shifts in the desirable direction (i.e., reduced political bias, reduced negative affect, increased journalistic writing quality), and bars extending from the right show the share shifting in the undesirable direction. Error bars indicate 95\% confidence intervals.}
\label{fig:avg}
\end{figure*}

Fig.~\ref{fig:avg} summarizes the main findings of our audit: relative to the source articles, AI news summarizers are broadly accurate, reduce political bias and negative affect, and improve journalistic writing quality along three of four dimensions. We elaborate on each finding below.

\subsection*{AI news summarizers are broadly accurate}

We evaluate the accuracy of AI-generated news summaries using a two-step process: first, we prompt an LLM to decompose each summary into decontextualized sentences; then, we ask it to assess whether each sentence is factually \textit{supported}, \textit{contradicted}, or \textit{unverifiable} given the source article (see~\ref{app:sec:data-annt} for prompts and human validation). This approach yields a quantitative measure of accuracy for each summary, computed as the percentage of factually supported sentences out of all extracted sentences. We also report the percentage of contradicted and unverifiable sentences to show the composition of inaccuracies.

First, as shown in Fig.~\ref{fig:avg}, AI-generated summaries achieve an average accuracy of 82.8\%, meaning that more than 8 out of 10 sentences in an AI-generated summary present facts that are fully supported by the source article. Only 2.5\% of sentences in an average AI summary contain a fact contradicted by the source article, and 14.7\% contain a fact that cannot be verified against it. A qualitative analysis of 90 summaries (880 sentences) detailed in \ref{app:sec:res-acc} further shows that inaccuracies rarely arise from entirely fabricated content, but primarily reflect linguistic imprecision in compressing or rephrasing content, or supplementary information added to the summary.

\begin{figure*}[h]
\centering
\includegraphics[width=\linewidth]{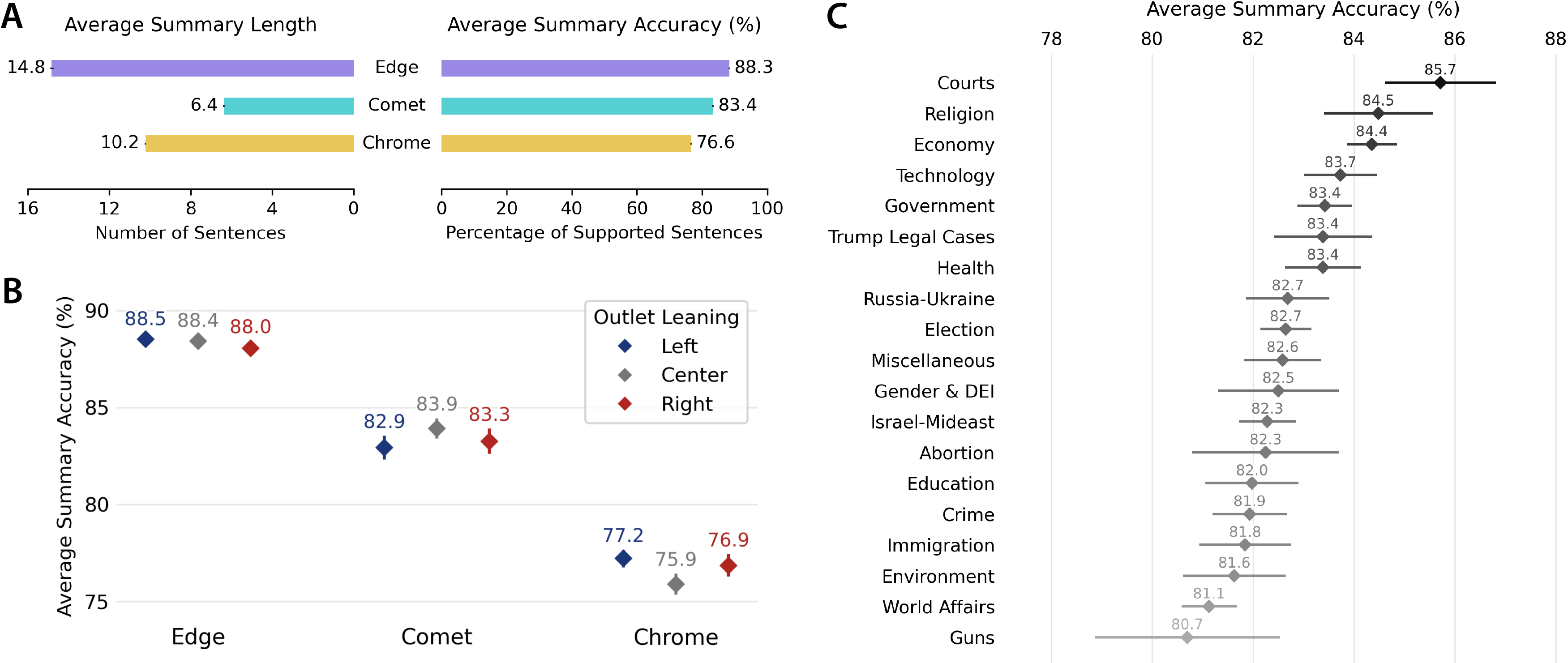}
\caption{Average summary accuracy by (A) browser, (B) outlet leaning, and (C) topic. Error bars indicate 95\% confidence intervals.}
\label{fig:acc}
\end{figure*}

Fig.~\ref{fig:acc}A shows that accuracy varies considerably between the three browsers tested. Edge is the most accurate (avg. 88.3\% supported, 1.8\% contradicted, 9.9\% unverifiable), Chrome the least (avg. 76.6\% supported, 3.3\% contradicted, 20.1\% unverifiable), and Comet falls in between (avg. 83.4\% supported, 2.4\% contradicted, 14.3\% unverifiable). In~\ref{app:sec:res-acc}, we show that these between-browser differences remain highly statistically significant (p<.001) after accounting for outlet leaning, topic, summary length, and article-level variation. Fig.~\ref{fig:acc-dist} additionally presents the accuracy distribution for each browser.

When it comes to variations by outlet's political ideology, Fig.~\ref{fig:acc}B plots summary accuracy for news articles from left-leaning, centrist, and right-leaning outlets. Clearly, the three AI summarization tools tested are similarly accurate across outlets of different leanings. That is, accuracy varies little within any tested browser and no outlet leaning is consistently favored across browsers. In Edge, news summary accuracy is nearly identical across left-leaning (88.5\%), centrist (88.4\%) and right-leaning (88.0\%) outlets, and although Comet (82.9\%/83.9\%/83.3\%, respectively) and Chrome (77.2\%/75.9\%/76.9\%, respectively) show slightly wider spreads, no difference across outlet leanings is statistically significant. In~\ref{app:sec:res-acc}, we report summary accuracy for individual news outlets.

We additionally test how summary accuracy varies across 19 salient political topics, ranging from the U.S. election and Trump's legal cases to contested social issues such as immigration, abortion, and gender, as well as international affairs such as the Israel–Middle East and Russia–Ukraine conflicts (see~\ref{app:sec:topic} for details of the topic model). Fig.~\ref{fig:acc}C shows that news summary accuracy varies only modestly across these widely diverse topics. 
As seen, summary accuracy is highest for news reporting on Courts (85.7\%), Religion (84.5\%), and Economy (84.4\%), and lowest for those on Guns (80.7\%), World Affairs (81.1\%), and Environment (81.6\%). Yet, these between-topic differences do not surpass 5\% and most are not statistically significant, as shown in~\ref{app:sec:res-acc}. 


In short, given prevailing concerns about LLM hallucinations, these results show that AI-generated news summaries are more accurate than widely feared and than previous manual audits suggested \cite{bbc2025ai, ebu2025news}. Across all three browsers, summaries are largely faithful to their source articles, and this faithfulness holds consistently across outlet leanings and topics.

\subsection*{AI news summarizers reduce political bias}

Does AI summarization alter the political bias of news articles? Our systematic audit assesses three dimensions of political bias: \textit{ideological bias}, or the political slant conveyed through the text's framing, tone, or evaluative language, on a five-point scale from far left to far right; as well as \textit{stance toward Democrats} and \textit{stance toward Republicans}, measured as the expressed favorability toward each party, its politicians, and policies, labeled as negative, neutral, or positive. Each article and summary is annotated using an extensively validated LLM (see~\ref{app:sec:data-annt} for prompts and validation).

Overall, AI news summarization tools reduce political bias across all three measures: summaries exhibit significantly less ideological bias and weaker partisan stances than their source articles (Wilcoxon signed-rank and McNemar, all p<.001, Holm-corrected). For each measure, Fig.~\ref{fig:avg}B illustrates the browser-averaged rates of bias \textit{reduction} in summaries of partisan articles as well as rates of bias \textit{introduction} in summaries of neutral articles. 

First, for ideological bias, 28.3\% of 2,919 left-leaning articles and 33.3\% of 2,477 right-leaning articles receive summaries with less ideological bias; meanwhile, 94.7\% of 8,381 neutral articles receive neutral summaries. In other words, more than a quarter of ideologically biased news is summarized in more neutral terms, and ideological neutrality is almost always preserved in summaries. 

\begin{figure*}[h]
\centering
\includegraphics[width=\linewidth]{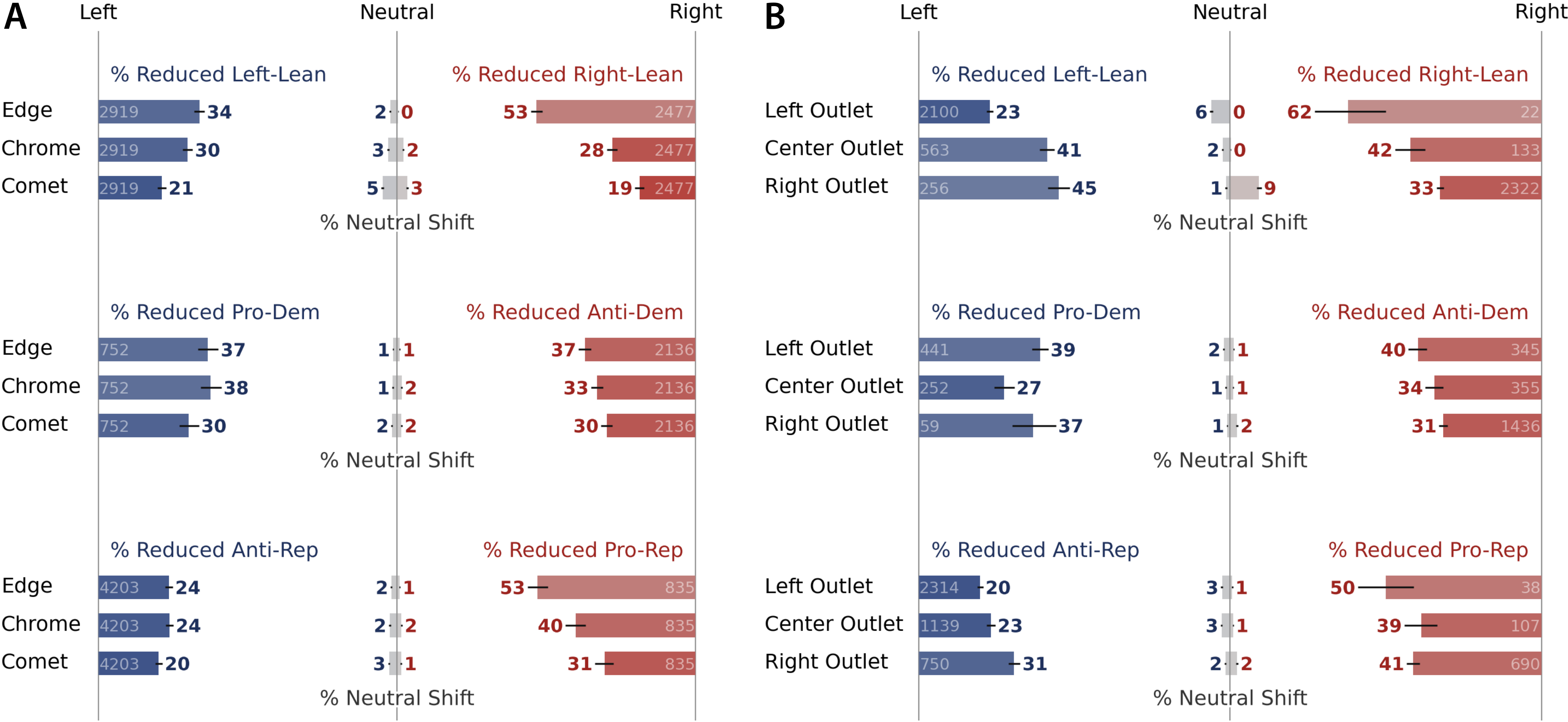}
\caption{Change in political bias (\textit{ideological bias}, \textit{stance toward Democrats}, and \textit{stance toward Republicans}) introduced by AI news summarizers, by (A) browser and (B) outlet leaning. In each panel, blue bars extending from the left show the share of left-leaning/pro-Democratic/anti-Republican articles receiving more neutral summaries, and red bars extending from the right show the share of right-leaning/anti-Democratic/pro-Republican articles receiving more neutral summaries. Gray bars in the middle show the share of neutral articles receiving left-leaning (toward the left) or right-leaning (toward the right) summaries. Small numbers within bars are article counts. Error bars indicate 95\% confidence intervals.}
\label{fig:main-ideo}
\end{figure*}

These aggregate patterns mask striking differences across browsers, illustrated in Fig.~\ref{fig:main-ideo}A. In particular, Edge appears to be aggressive in reducing ideological bias for right-leaning articles. It does so far more often than other browsers: 53\% of right-leaning articles receive more neutral summaries in Edge, compared to only 28\% in Chrome and 19\% in Comet (both p<.001). In addition, Edge reduces ideological bias for right-leaning articles far more often than for left-leaning articles (right: 53\% vs. left: 34\%, p<.001), an asymmetry that is absent in both Chrome (right: 28\% vs. left: 30\%, p=.37) and Comet (right: 19\% vs. left: 21\%, p=.43).
\ref{app:sec:res-ideo} details statistical tests that assess the effects of browser, outlet leaning, and topic on bias reduction, and presents a qualitative analysis to explain the distinctive behavior of Edge. In short, we show that AI summarization tools consistently reduce ideological bias, though with notable differences between tools.

Similar overall patterns emerge for stances toward the two major parties. AI summarizers substantially decrease both favorability and hostility for a large proportion of partisan articles, while preserving neutral stances in over 95\% of cases. However, the specific patterns differ sharply between the two parties. For stances toward Democrats, on average, 34.9\% of 752 pro-Democratic articles and 33.3\% of 2,136 anti-Democratic articles receive summaries with a more neutral stance.
This symmetry holds across browsers (pro/anti: 37\%/37\% in Edge, 38\%/33\% in Chrome, 30\%/30\% in Comet). 
In contrast, AI summarizers, and especially Edge, reduce political bias much more aggressively for news articles favorable to the Republicans. On average, 41.2\% of 835 pro-Republican articles receive summaries 
that temper the source article's positivity. In turn, only 22.6\% of 4,203 anti-Republican articles receive more neutral summaries. This asymmetry holds across all browsers (pro/anti: 53\%/24\% in Edge, 40\%/24\% in Chrome, 31\%/20\% in Comet), with all differences highly significant (p<.001). 

Looking across parties, we find that anti-Democratic stances are significantly more likely to be attenuated than anti-Republican stances by all three browsers (all p<.001), and pro-Republican stances are more likely to be attenuated than pro-Democratic stances (p<.001 for Edge). Although this asymmetry raises concerns about bias in AI summarizers, a qualitative analysis detailed in~\ref{app:sec:res-ideo} suggests that the pattern may simply reflect how partisan stances are expressed differently in the source articles.

Furthermore, as shown in Fig.~\ref{fig:main-ideo}B, the observed political bias reduction---for both ideological bias (top panel) and partisan stances (two bottom panels)---holds across outlet leanings. Whether the news article is published in a left, centrist, or right outlet, AI summarizers reduce political bias much more often than they increase it, with pro-Republican stances attenuated more often than anti-Republican ones. Although the magnitude of these effects does vary across outlet leanings---as we discuss in~\ref{app:sec:res-ideo}---
the broader pattern of political bias reduction persists despite these variations.


This pronounced reduction of political bias in AI news summaries is also consistent across all topics. As shown in~\ref{app:sec:res-ideo}, across measures of \textit{ideological bias}, \textit{stance toward Democrats}, and \textit{stance toward Republicans}, political bias is reduced more often than it is introduced for every topic, with pro-Republican articles consistently receiving neutral summaries more often than anti-Republican articles. 

\subsection*{AI news summarizers reduce negative affect}

Next, we examine how AI summarization alters three dimensions of negative affect: \textit{negativity}, whether the language and framing of the text are negative or not; \textit{anger}, the intensity of frustration, irritation, or rage expressed, coded as strong, moderate, or absent; and \textit{fear}, the intensity of anxiety, apprehension, or dread, coded as strong, moderate, or absent. For each article and summary, negativity is annotated using a transformer-based classifier, while anger and fear are annotated using an LLM (see~\ref{app:sec:data-annt} for classifier details, prompts, and validation).

\begin{figure*}[h]
\centering
\includegraphics[width=\linewidth]{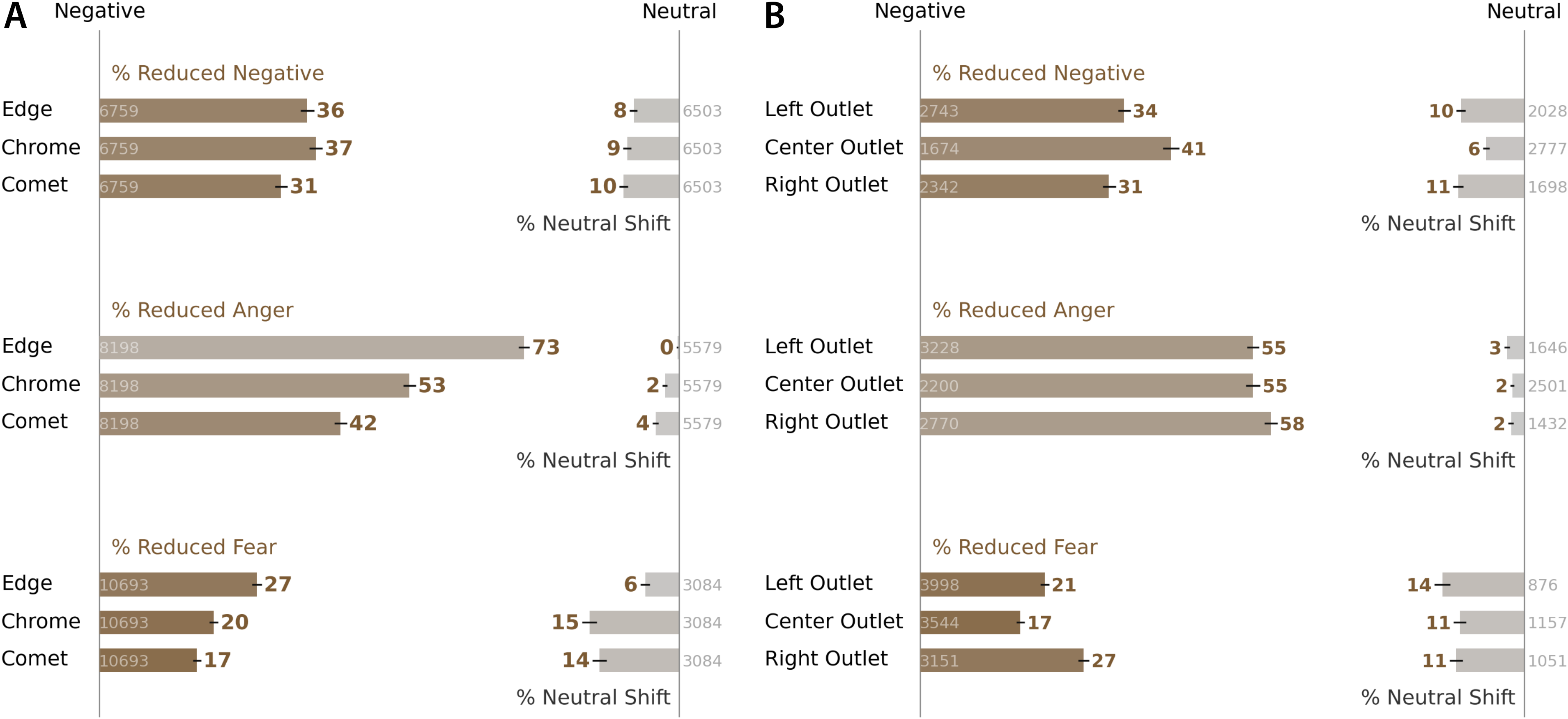}
\caption{Change in negative affect (\textit{negativity}, \textit{anger}, and \textit{fear}) introduced by AI news summarizers, by (A) browser and (B) outlet leaning. In each panel, bars extending from the left show the share of articles with negativity/anger/fear receiving more neutral summaries, and bars extending from the right show the share of articles without negativity/anger/fear receiving summaries with more negative affect. Small numbers within or besides bars are article counts. Error bars indicate 95\% confidence intervals.}
\label{fig:main-neg}
\end{figure*}

Across all three measures, AI news summaries exhibit significantly less negative affect than their source articles (McNemar and Wilcoxon signed-rank, all p<.001, Holm-corrected). For negativity, 34.8\% of negative articles receive more neutral summaries, while 8.8\% of neutral articles receive more negative summaries, averaged across the three browsers. 
Fig.~\ref{fig:main-neg}A illustrates the rates of article-to-summary transitions, broken down by browser.
Differences across browsers are modest but statistically significant (see~\ref{app:sec:res-emo} for details of the tests): compared to Chrome, which reduces negativity for 37\% of negative articles, Comet does so for substantially fewer (31\%, p<.001), and Edge for marginally fewer (36\%, p<.05).

The reduction of anger is even more pronounced. AI summarizers tone down anger for the majority (56.1\%) of articles expressing anger, and almost never (2.3\%) introduce anger into the summaries of articles without it. We find sharp divergence between the browsers: compared to Chrome, which reduces anger for 53\% of articles that express it, Edge does so significantly more often (73\%, p<.001), and Comet significantly less (42\%, p<.001).

In contrast, AI summarizers reduce fear less often and introduce it more, although fear is still reduced more than increased. On average, 21.2\% of articles expressing fear receive summaries with less fear, and 11.7\% of no-fear articles receive summaries with fear. The fear reduction and introduction rates are especially close in Comet and Chrome (reduced/introduced: 17\%/14\% in Comet, 20\%/15\% in Chrome, 27\%/6\% in Edge). In~\ref{app:sec:res-emo}, we present a qualitative analysis of summaries with increased negative affect, to probe the causes and implications of such cases.

In short, as with political bias, AI summarizers reduce negative affect consistently more often than they introduce it, with anger reduced most and fear reduced least. As seen in Fig.~\ref{fig:main-neg}B, this pattern holds across all outlet leanings, despite modest variations discussed in~\ref{app:sec:res-emo}. 
Furthermore, we see similar negative affect reduction across all topics, as shown in Fig.~\ref{fig:topic}. 

\subsection*{AI news summarizers enhance journalistic writing quality}

Lastly, we evaluate how AI summarization changes 
four aspects of journalistic writing: \textit{clarity}, whether key ideas of the text are immediately clear and easy to follow; \textit{engagement}, whether the writing is vivid and attention-grabbing; \textit{personal tone}, whether the text foregrounds the author's own opinions or human-interest elements rather than the facts; as well as \textit{clickbait and sensationalism}, whether the text is sensationalistic, unnecessarily repetitive, or misleadingly titled. Each article and summary is annotated using a transformer-based classifier we developed (see ~\ref{app:sec:data-annt} for full details on the architecture, training, and validation). 

Fig.~\ref{fig:avg}D
shows that AI-generated news summaries largely exhibit enhanced writing quality relative to their source articles. The numbers are striking: on average, 80.3\% of low-clarity articles receive summaries that are clearer and easier to follow, 72.7\% of articles with a personal tone receive more objective summaries, and 86.6\% of articles which use clickbait tactics or sensationalism receive summaries without any; meanwhile, only 2.2\%, 1.7\%, and 0.2\% of articles receive summaries with lower clarity, a more personal tone, and more clickbait tendencies, respectively. In effect, AI summarizers overwhelmingly increase clarity, temper personal tone, and strip clickbait, while almost never doing the reverse. As illustrated in Fig.~\ref{fig:main-quality}, browser differences are consistent across dimensions yet modest in magnitude: Comet most often produces summaries of higher writing quality (for 83\% of low-clarity articles, 76\% with a personal tone, and 89\% using clickbait tactics), Edge least (76\%/64\%/85\%), and Chrome falls in between (81\%/74\%/85\%).

The only exception is the change in engagement. Perhaps not surprisingly, AI summarizers tend to flatten engaging language present in source articles by adopting a formulaic summary structure and a dense, detached, and information-forward writing style. Therefore, the majority of articles receive medium-engagement summaries, regardless of their source articles' engagement level. Quantitatively, while 39.9\% of low-engagement articles receive more engaging summaries (Edge/Chrome/Comet: 32\%/40\%/48\%), 78.0\% of high-engagement articles receive less engaging summaries (Edge/Chrome/Comet: 82\%/81\%/71\%). As high-engagement articles substantially outnumber low-engagement ones (2,305 vs. 242), AI-powered browsers on average produce less engaging summaries. Statistical tests confirm the overall significant increase in clarity and decrease in personal tone, clickbait tactics, and engagement (McNemar and Wilcoxon signed-rank, all p<.001, Holm-corrected). As shown in Fig.~\ref{fig:main-quality} and~\ref{fig:qual-topic}, the patterns also hold across all outlet leanings and topics, although the magnitude of the effects varies.

\section*{Discussion}\label{{sec:discussion}}
AI-powered browsers are emerging as a new layer of information intermediation between publishers and audiences \cite{google2025go, microsoft2025meet}. 
By automatically generating summaries of news articles, these systems increasingly determine what information users encounter, how it is framed, and which aspects of a story receive emphasis. Yet, despite their growing role in the online news ecosystem, little is known about how browser-integrated AI systems transform news content in practice. Auditing three major AI-powered browsers across 41,331 summaries of 13,777 news articles from 15 outlets spanning the ideological spectrum, we find that AI summarization reshapes the informational character of news in ways that extend beyond simple compression.

Our first key finding is that browser-based AI summarization is generally accurate. Roughly 83\% of sentences in AI-generated news summaries are supported by their corresponding source articles, with Microsoft Edge being the most accurate (88\%). Of course, these tools are far from error-free: across browsers, approximately one in six summary sentences is not fully supported by the source article. However, our qualitative analysis suggests that most inaccuracies are relatively minor. As we show in~\ref{app:sec:res-acc}, they primarily stem from imprecise wording in summarization or from added contextual information that tends to be factually accurate. An estimated 5\% of inaccuracies (that is, less than 1\% of all sentences) involve made-up claims that are neither supported nor implied by the source; and less than 3\% of sentences in summaries contradict their source articles, with many contradictions being small errors, such as confusing a plural for a singular when attributing a fact. This finding revises the pessimistic picture of AI-mediated news consumption drawn by previous large-scale audits. 
Although these audits suggest concerning inaccuracies when AI chatbots search for and synthesize sources to answer questions about news \cite{bbc2025ai, ebu2025news}, we show that browser-based summarization of individual articles---a rapidly spreading and arguably more consequential mode of AI-mediated news consumption---renders news content with promising accuracy.


Second, we find that AI summarization systematically attenuates partisan and emotional content in news. Compared to source articles, AI-generated summaries are substantially less ideologically biased, have less pronounced partisan stances, lower negativity, and reduced expressions of anger and fear. Our results extend existing evidence that LLM summarization reduces the ideological bias of news content in controlled chatbot evaluations \cite{savgira2026stays}, by demonstrating that these patterns also emerge in commercially deployed browser environments that increasingly mediate everyday news consumption \cite{newman2025reuters}. Furthermore, we show that the observed attenuation extends beyond ideological bias to partisan stances and various dimensions of negative affect. Across all indicators, AI summarization reduces partisan bias and negative affect far more often than it introduces them, with reduction rates spanning 21.2--56.1\% (avg. 34.0\%) and introduction rates 1.3--11.7\% (avg. 3.8\%). Despite notable asymmetries in the degree of attenuation across measures---for example, a pro-Republican stance is attenuated more often than an anti-Republican one, and anger more often than fear---our audit provides strong evidence that AI summarization systematically compresses multiple dimensions of bias and negativity in news content.

Third, AI summarizers present news with largely enhanced journalistic writing quality. Relative to their source articles, AI news summaries are clearer and easier to follow, more restrained on personal opinions and anecdotes, and less reliant on clickbait tactics and sensationalism. These positive shifts likely stem from the impersonal, information-forward style that LLMs adopt in general \cite{munoz2024contrasting,reinhart2025llms,markey2024dense}, although the same style also renders AI summaries less engaging to read, as vivid storytelling gets flattened into a dense, detached format.

Lastly, all three patterns---accurate summarization, attenuation of political bias and negative affect, and general improvement in journalistic writing quality---are remarkably consistent. Across three independent browser ecosystems, left-leaning, centrist, and right-leaning news outlets, and 19 substantive news topics, we observe broadly similar patterns of accuracy, attenuated bias and negativity, and largely enhanced writing quality. 
Although there are meaningful variations in the magnitude of effects---for example, Edge attenuates pro-Republican stances more aggressively than other browsers---the dominant patterns 
hold regardless of the browser tested, the ideological leaning of the source outlet, or the topical domain of the news article.
This finding contributes to ongoing debates about algorithmic fairness by suggesting that contemporary browser-based summarization systems apply transformations in consistent directions across politically diverse sources and numerous distinct issues, even as their magnitude varies.

Taken together, our findings highlight the ways in which AI-powered browsers function 
as editorial intermediaries. 
In the process of condensing news, these browsers systematically attenuate its ideological intensity, partisan stances, and negative affect, while also increasing clarity and readability. This may benefit news consumers by minimizing cognitive burden, making news content easier to digest, and reducing ideological bias, explicit support or opposition toward political parties, and the negativity, fear, and anger that are increasingly central to news reporting \cite{rozado2022longitudinal, semetko2000framing}. On the one hand, these effects may be seen as democratically beneficial. After all, biased and highly partisan news generates polarization \cite{levendusky2013partisan} and outparty hostility \cite{levendusky2013partisan2}, and negative reporting decreases well-being \cite{boukes2017news} and turns many people away from news and politics \cite{wertz2026correlates, villi2022taking}.
On the other hand, however, AI summarization tools may alter key features of political discourse. Journalistic choices about framing, emphasis, emotional tone, and evaluative language are not incidental aspects of reporting; they often convey information about the importance, urgency, and political meaning of events and, some argue, serve to render journalists' own biases and perspectives more transparent \cite{hanitzschMappingJournalismCultures2011}. By systematically attenuating these features, AI summarization tools present users with a version of the news that is less reflective of the original intent of journalists and publishers, 
and may make 
news less engaging to some readers. 


Our work opens several key directions for future research. First, our analysis focuses on three browser-based summarization systems during a specific time period. Because the underlying models and product designs are proprietary and subject to continuous revision, the observed effects may change as these systems evolve. Longitudinal audits will be necessary to assess the stability of outputs generated by AI news summarization tools over time. Second, although our dataset spans a large number of outlets and topics, it is limited to English-language news and the U.S. political context. Future research should examine whether similar patterns emerge in other countries, languages, and political or media systems. Third, our audit measures changes in content rather than their effects on audiences. Whether the observed reductions in ideological and emotional intensity influence polarization, political attitudes, perceptions of bias, trust in journalism, or news engagement remains an open question. Experimental studies examining how users respond to AI-generated summaries relative to original articles are a needed next step. Finally, although our audit identifies systematic algorithmic attenuation of various features of news content, it cannot directly determine the mechanisms that produce these patterns. Building on our work, future audits should test how model architectures, system prompts, safety policies, and interface design choices jointly shape the effects documented here: summarization accuracy, attenuation of partisan and emotional content, and improvement in journalistic writing quality. 


As browser-based AI summarization tools become increasingly integrated into everyday information environments, they are likely to play a growing role in shaping political learning, opinion formation, trust in news, and democracy more broadly. Our findings show that these tools do not simply condense the news; they systematically transform it. Understanding the societal consequences of this new form of algorithmic mediation represents an important challenge for future multidisciplinary research.  

\section*{Materials and Methods}\label{sec:methods}

\subsection*{Study design}

We audit the news-summarization features built into three widely used AI-powered browsers---Perplexity Comet, Microsoft Edge (summaries produced by Copilot), and Google Chrome (summaries produced by Gemini)---as end users encounter them. Rather than approximating these features by submitting article text to LLMs with a custom prompt, we employ an automated pipeline that opens each live news page in the target browser, invokes the browser's default summarization function, and saves the returned summary, processing each article in a fresh conversation to prevent context carryover. The unit of analysis is the article-summary pair, and our target of inference is the behavior of each deployed browser-and-model system during the collection window, not the properties of the underlying language models in isolation.

\subsection*{Data collection}

We assembled a corpus of political news articles published between 1~January 2024 and 30~June 2025 from 15 U.S.\ outlets spanning the political spectrum (five left-leaning, five centrist, five right-leaning). For each outlet, we randomly sampled 1{,}000 articles whose headlines were classified as political by a transformer-based classifier \cite{talaga2025political}. We obtained article texts from Media Cloud \cite{roberts2021media}, except for 6 paywalled outlets where we scraped the articles directly using authenticated subscriptions.
Summary collection ran on nine parallel virtual machines routed through a single U.S.-based IP address across waves in February and March 2026, with later waves used to re-attempt failed collections.
Invalid summaries were removed using LLM-assisted screening.
To ensure that cross-browser comparisons rested on the same underlying articles, we restricted the analysis to articles for which all three browsers returned a valid summary. Then, we removed pure video or podcast items and articles shorter than 100~words, yielding a common analytical sample of 13,777 articles, each represented by three summaries. \ref{app:sec:data-col} provides more details on outlet selection, collection setup, and screening procedure.

\subsection*{Measurement and validation}

A variety of machine classifiers were used to identify differences between source articles and their corresponding summaries. To ensure comparable measurements, the same tools were used to assess both articles and summaries.
The specific measurements are described below. 

\textit{Ideological bias} was measured using GPT-5.2, on an ordinal scale from -2 (very left) to 2 (very right).

\textit{Stances toward Democrats} and \textit{toward Republicans} were also measured using GPT-5.2, each on an ordinal scale of -1 (negative), 0 (neutral), and 1 (positive). 

\textit{Negativity} was measured using a previously developed transformer-based classifier designed to identify negative sentiment in news, which outputs 0 (non-negative) or 1 (negative).

\textit{Fear} and \textit{anger} were both measured using GPT 5.2, on an ordinal scale of 0 (none), 1 (moderate), and 2 (strong).

\textit{Clarity}, \textit{engagement}, \textit{personal tone} and \textit{clickbait} were measured using a previously developed transformer-based classifier. Clarity was scored 0 (unclear) or 1 (clear), and all other measures used a scale from 0 (unengaging/impersonal/no-clickbait) to 2 (extremely engaging/personal/sensationalistic).

In addition, the \textit{accuracy} of each summary was assessed using GPT-5.5 through a two-step process: the summary was first decomposed into sentences, and each sentence was then labeled supported, contradicted, or unverifiable against the source article; accuracy was then calculated as the proportion of sentences labeled supported. 

We extensively validated all above measures against human annotation. On a sample of 880 sentences from 90 summaries, GPT-5.5's accuracy labels achieved 94\% agreement with human labels and an $F_1$ score of .91. On a sample of 25 articles and their 75 summaries, automated annotations reached an accuracy of 80\% on \textit{ideological bias}, 92\% on \textit{stance toward Democrats}, 82\% on \textit{stance toward Republicans}, 85\% on \textit{negativity}, 80\% on \textit{anger}, 74\% on \textit{fear}, 95\% on \textit{clarity}, 89\% on \textit{engagement}, 90\% on \textit{personal tone}, and 99\% on \textit{clickbait}.

Additional details regarding LLM prompts, classifiers, and validation procedures are reported in~\ref{app:sec:data-annt}.

\subsection*{Statistical analysis}

To test for differences between article and summary scores, we performed Wilcoxon signed-rank tests for all ordinal measures and McNemar tests for all binary measures, with each article-summary pair as the unit of analysis. We applied Holm's correction for multiple comparisons across measures of the same construct. 

Furthermore, we used mixed-effects regression models to examine the effects of browser, outlet leaning, and topic on each outcome of interest. For accuracy, we fit an ordered beta regression model \cite{kubinec2023ordered}. For each measure of partisan bias, negative affect, and journalistic writing quality, we fit a logistic regression model, with outcome being \textit{reduced} for partisan bias and negative affect---whether the summary exhibits reduced partisan bias or negative affect---and \textit{increased} for journalistic quality. We included browser, outlet leaning, and topic as categorical fixed effects, standardized article length as a control, and random intercepts for articles nested within outlets. All models were fit using the \texttt{glmmTMB} \cite{brooks2017glmmtmb} package in \texttt{R}. Full details of the regression models and results are provided in~\ref{app:sec:res}.


\bibliographystyle{elsarticle-num} 
\bibliography{references}

{
\small

\section*{Funding}

The authors gratefully acknowledge the support of the European Research Council (ERC Consolidator 101126218, NEWSUSE: Incentivizing Citizen Exposure to Quality News Online: Framework and Tools, PI Magdalena Wojcieszak) and of the Center for Excellence in Social Sciences at the University of Warsaw, which funded the Red Giant funding project (PI Magdalena Wojcieszak). Any opinions, findings, conclusions, or recommendations expressed in this material are those of the authors and do not necessarily reflect the views of the European Research Council or the University of Warsaw.

\section*{Acknowledgments}\label{sec:acknowledgments}
This research was carried out with the support of the Interdisciplinary Centre for Mathematical and Computational Modelling, University of Warsaw (ICM UW), under computational allocation no. G104-2654. We also thank Joshua Mariano for data annotation.

\subsection*{Author contributions}
D.B., E.W., and M.W. designed the research. Y.X., D.B., and M.M. oversaw the data collection. Y.X., E.W., L.L., and M.W. analyzed the data. Y.X., D.B., E.W., and M.W. wrote the manuscript. All authors revised the manuscript.

\subsection*{Competing interests}
The authors declare that they have no competing interests.


\subsection*{Data and materials availability}
All data needed to evaluate the conclusions in the paper are present in the paper, the Supplementary Materials, or at Zenodo (\url{https://doi.org/10.5281/zenodo.21457591}).

}

\clearpage

\appendix
\setcounter{table}{0}
\onecolumn

\section{Data}\label{app:sec:data}

\subsection{Data Collection}
\label{app:sec:data-col}

\paragraph{News outlets and source corpus}
We selected 15 U.S.\ news outlets to span the political spectrum, with five outlets in each of three leaning categories (left, center, right). Outlet leaning was assigned using the bias ratings of Media Bias/Fact Check (MBFC), applying a cut-off of $\pm 3.5$ on the MBFC scale: outlets scoring below $-3.5$ were classified as left-leaning, those above $+3.5$ as right-leaning, and the remainder as centrist. Table~\ref{tab:outlets} lists the outlets along with their leaning, MBFC bias rating, journalistic quality score rated by a prior study \cite{lin2023high}, paywall status, and number of articles in the final data set.

\begin{table}[ht]
\centering
\caption{News outlets, political leaning, bias and quality ratings, paywall status, and number of articles in the final data set. Bias ratings are from Media Bias/Fact Check (negative = left, positive = right; classification cut-off $\pm 3.5$). Quality scores are from Lin et al. \cite{lin2023high}. A ``hard'' paywall indicates that full article text is inaccessible without a subscription, and a``soft'' one indicates that access is metered or partially restricted.}
\label{tab:outlets}
\begin{tabular}{llrrlr}
\toprule
Outlet & Leaning & MBFC Bias & Quality & Paywall & \#Articles \\
\midrule
MSNBC                   & Left   & $-6.4$ & 0.59 & & 996 \\
The New York Times      & Left   & $-4.1$ & 0.86 & Yes (hard) & 892 \\
CNN                     & Left   & $-3.6$ & 0.66 & & 986 \\
NBC News                & Left   & $-3.6$ & 0.84 & & 1{,}000 \\
The Guardian            & Left   & $-3.6$ & 0.75 & & 1{,}000 \\
Associated Press        & Center & $-2.1$ & 1.00 & & 1{,}000 \\
Reuters                 & Center & $-0.5$ & 1.00 & Yes (soft) & 924 \\
The Hill                & Center & $0.4$  & 0.90 & & 979 \\
Forbes                  & Center & $1.3$  & 0.83 & Yes (soft) & 801 \\
CBS News                & Center & $2.2$  & 0.88 & & 997 \\
Reason                  & Right  & $3.8$  & 0.81 & & 875 \\
The Free Press          & Right  & $4.1$  & ---  & Yes (soft) & 984 \\
The Wall Street Journal & Right  & $4.2$  & 0.80 & Yes (hard) & 686 \\
OANN                    & Right  & $7.9$  & 0.41 & & 664 \\
Fox News                & Right  & $8.0$  & 0.53 & & 993 \\
\bottomrule
\end{tabular}
\end{table}

For each outlet we compiled 1{,}000 articles published between 1~January 2024 and 30~June 2025. Candidate articles were restricted to political content: every article headline was classified by the ERC-NEWSUSE political content classifier \cite{talaga2025political}---a fine-tuned transformer model---and only articles with headlines classified as political were retained. For nine outlets, the full article texts were available from Media Cloud \cite{roberts2021media}; for the remaining six outlets, the article texts were obtained by direct scraping. For outlets operating behind a paywall, summarization was performed using authenticated subscriptions so that the browser could access the complete article text rather than a truncated preview.

Two outlets initially considered for inclusion, Breitbart and ZeroHedge, were excluded because the browsers consistently declined to summarize their pages even though the underlying articles were accessible; for example, Comet responded to Breitbart pages with a refusal stating that the page could not be accessed due to the user's or organisation's settings. Because such refusals were systematic rather than article-specific, these outlets were dropped prior to data collection.

\paragraph{Browsers and summarization features}
We audited the default, user-facing summarization feature of each browser as deployed to ordinary users. Chrome and Comet were used without a subscription, relying on its default model with no model selection options. Edge was used with an institutional subscription enabling higher Copilot usage limits. In each case, the summary was produced by invoking the browser's built-in assistant on the open article page, and requesting a summary using the assistant's default summarization action (Comet) or its default summarization prompt (Chrome, Edge). The exact browser builds used in each collection wave are listed in Table~\ref{tab:rounds}.

\paragraph{Collection procedure and infrastructure}

Summaries were collected through an automated pipeline that emulated ordinary user interaction with each browser. For each article URL, the script opened the page in the target browser, invoked the built-in AI assistant, requested a summary of the open page using the assistant's default function, copied the generated summary, and saved the returned text. Each article was processed in a new conversation so that no context carried over between summaries. To collect data at scale, we ran nine virtual machines in parallel, all routed through a single U.S.-based IP address; the IP location was held constant across browsers and waves so that any geolocation-dependent behavior of the summarization features was held fixed. Data were collected in successive waves during February and March 2026, with later waves used to re-attempt articles for which an earlier wave had not returned a valid summary. Table~\ref{tab:rounds} reports the collection waves, periods, and browser builds.

\begin{table}[ht]
\centering
\caption{Collection waves, periods, and browser builds.}
\label{tab:rounds}
\begin{tabular}{llll}
\toprule
Wave & Browser & Period & Browser build \\
\midrule
Wave 1 & Edge  & 16--28 Feb 2026 & 145.0.3800.65 \\
Wave 1 & Comet & 16--28 Feb 2026 & 144.0.7559.97 \\
Wave 1 & Chrome & 6--16 Mar 2026 & 145.0.7632.160 \\
\addlinespace[3pt]
Wave 2 & Edge  & 9--13 Mar 2026 & 145.0.3800.97 \\
Wave 2 & Comet & 9--13 Mar 2026 & 145.0.7632.76 \\
Wave 2 & Chrome & 19 Mar 2026 & 146.0.7680.154 \\
\addlinespace[3pt]
Wave 3 & Edge  & 18--19 Mar 2026 & 145.0.3800.97 \\
Wave 3 & Comet & 19 Mar 2026 & 145.0.7632.76 \\
\bottomrule
\end{tabular}
\end{table}

Not every collection attempt yielded a valid summary; three distinct failure modes occurred. First, an attempt sometimes did not produce a copied summary at all, due to rate limits or technical issues (e.g., summary generation running abnormally long and not finishing before the copy attempt). Second, the browser assistant sometimes generated text that was not a valid article summary, when the page was not accessible (e.g., a CAPTCHA appeared) or it contained no article text to summarize (e.g., only a video player and related links). Third, mostly in Chrome, the assistant sometimes produced a seemingly valid summary even when the page was blocked by a CAPTCHA.


To exclude invalid responses, we ran each collected summary through a three-step filter. First, we removed all empty summaries resulting from failed copy attempts. Second, we used GPT-5.2-2025-12-11 to classify whether each summary is a valid article summary or not; only summaries classified as valid were retained. Third, to filter out seemingly valid summaries of CAPTCHA-blocked pages, we captured a screenshot of each page during summarization, and discarded summaries where the screen text contained known CAPTCHA phrases (e.g., ``verification required'', ``confirm that you are human'', ``slide right to secure''). Table~\ref{tab:attrition} reports the number of articles that remained after each filtering step, for each browser and collection wave.


Since some failure modes were concentrated in particular browsers, the three browsers initially yielded overlapping but non-identical sets of successfully summarized articles. 
To ensure that all cross-browser comparisons rested on the same underlying articles, we restricted the analysis to the set of 14{,}365 articles for which all three browsers returned a valid summary. From this common set, we additionally removed 185 articles shorter than 100~words, which are too brief to summarize meaningfully, and 403 articles consisting purely of video or podcast content. This produced the final balanced panel of 13{,}777 articles, each represented by three summaries (one per browser).

\begin{table}[ht]
\centering
\caption{Number of remaining articles after each filtering step, by browser and collection wave. ``Captured'' counts responses retrieved; ``Summaries'' counts
responses passing the GPT-5.2 summary screening; ``Retained'' counts those remaining after removing CAPTCHA-affected responses. Bold totals are carried forward to the three-browser intersection.}
\label{tab:attrition}
\small
\begin{tabular}{lrrrr}
\toprule
 & Attempted & Captured & Summaries & Retained \\
\midrule
\emph{Edge --- Wave 1} \\
 & 15{,}000 & 9{,}777 & 9{,}554 & 9{,}552 \\
\multicolumn{5}{l}{\emph{Edge --- Wave 2}} \\
 & 5{,}447 & 4{,}598 & 4{,}533 & 4{,}528 \\
\multicolumn{5}{l}{\emph{Edge --- Wave 3}} \\
 & 919 & 527 & 495 & 495 \\
\addlinespace[3pt]
\multicolumn{4}{l}{\emph{Edge --- Total}} & \textbf{14{,}575} \\
\midrule
\multicolumn{5}{l}{\emph{Comet --- Wave 1}} \\
 & 15{,}000 & 14{,}180 & 12{,}635 & 12{,}628 \\
\multicolumn{5}{l}{\emph{Comet --- Wave 2}} \\
 & 2{,}371 & 1{,}995 & 1{,}933 & 1{,}933 \\
 \multicolumn{5}{l}{\emph{Comet --- Wave 2}} \\
 & 438 & 435 & 412 & 412 \\
\addlinespace[3pt]
\multicolumn{4}{l}{\emph{Comet --- Total}} & \textbf{14{,}973} \\
\midrule
\multicolumn{5}{l}{\emph{Chrome --- Wave 1}} \\
 & 15{,}000 & 14{,}353 & 14{,}212 & 13{,}728 \\
\multicolumn{5}{l}{\emph{Chrome --- Wave 2}} \\
 & 1{,}271 & 1{,}241 & 1{,}216 & 1{,}072 \\
\addlinespace[3pt]
\multicolumn{4}{l}{\emph{Chrome --- Total}} & \textbf{14{,}800} \\
\bottomrule
\end{tabular}
\end{table}

\subsection{Data Annotation}
\label{app:sec:data-annt}
\paragraph{Accuracy}
We annotated the factual accuracy of each news summary against its source article using GPT-5.5-2026-04-23 through a two-step process. First, we used the prompt in Figure~\ref{fig:prompt_decomp} to decompose each summary into a list of decontexualized sentences. Then, we used the prompt in Figure~\ref{fig:prompt_eval} to evaluate if each sentence is factually \textit{supported}, \textit{contradicted}, or \textit{unverifiable} given the source article. We also reserved an \textit{invalid} label for sentences that contain no facts other than article metadata.

We manually validated the outputs of the two steps separately. For sentence decomposition, one annotator evaluated 206 decomposed sentences from 24 randomly sampled summaries, judging 99\% (204/206) as accurate. For accuracy evaluation, we validated the LLM outputs on a random sample of 880 sentences from 90 summaries (corresponding to 30 news articles, two from each outlet). Two annotators independently labeled a subsample of 269 sentences from 30 summaries, reaching 87\% observed agreement (Cohen's $\kappa$=.67). After adjudicating the disagreements, one of the annotators labeled the remaining sentences. The LLM's 880 accuracy labels achieved 94\% agreement with human labels and an $F_1$ score of .91.

\begin{figure}[h]
\centering
\fbox{
\begin{minipage}{0.8\linewidth}  
\small
Your task is to convert a news summary into a list of decontextualized sentences to be used for accuracy checks against its source article.

*Pre-Processing*

Remove:

- Preamble and postamble (e.g., greetings, offers to help)

- Formatting markers (e.g., bold, italics, bullets, emoji)

- Embedded urls

- Descriptions of information irrelevant to the main article content (e.g., user comments, related links, promotional text)

*Conversion Rules*

- If the input contains complete sentences, preserve these sentences and their current boundaries. Do not split or merge sentences.

- Convert non-sentence fragments (e.g., bullet points) into complete sentences while preserving the source wording. Merge fragments that are components of the same sentence.

- Make each sentence fully self-contained: a reader who has never seen the summary must be able to understand the sentence without any other context. Replace all pronouns, demonstratives, and implicit references with their explicit referents. For nested items, include relevant context from levels above.

- Preserve attribution, hedging, and specifics (e.g., names, dates, numbers, titles, places).

- Do not infer, expand, or add context from your own knowledge.

*Output Format*

Return the results as a numbered list of sentences. Do not include any explanations, comments, markdown, or code fences.

Example:

1. First sentence.

2. Second sentence.

3. Third sentence.

*Input*

The summary you have to convert:

\{summary\}
\end{minipage}
}
\caption{Prompt for decomposing summaries into sentences.}
\label{fig:prompt_decomp}
\end{figure}

\begin{figure}[h]
\centering
\fbox{
\begin{minipage}{0.8\linewidth}  
\small
Your task is to evaluate the *factual accuracy* of sentences from a news summary against the original source article. 

*Pre-Processing*

For each sentence, remove any references to article metadata (author, type, format, outlet, publication date, podcast name/series/host, etc.) or webpage metadata (links to related articles, embedded media, etc.). Label the sentence based only on the remaining facts that are part of the article content.

*Labels*

For each pre-processed sentence, choose one of the following labels:

- SUPPORTED: Every fact in the sentence is supported by the original article.

- CONTRADICTED: One or more facts in the sentence are contradicted by the article.

- UNVERIFIABLE: One or more facts in the sentence cannot be confirmed or denied based on the article. 

- INVALID: The sentence contains no facts other than metadata.

If both contradicted and unverifiable facts are present, label CONTRADICTED.

*Output*

Return the results as a numbered list in the following format. Include brief reasoning if the sentence is not labeled SUPPORTED.

1. Label: ...

   Reasoning: ...
   
2. ...

*Input*

Article:

\{title\} (from \{domain\}, published \{date\})

\{article\}

The sentences you have to label:

\{sentences\}
\end{minipage}
}
\caption{Prompt for evaluating the accuracy of summary sentences.}
\label{fig:prompt_eval}
\end{figure}

\paragraph{Political bias}
We used GPT-5.2-2025-12-11 to annotate each news article and summary for \textit{ideological bias}, \textit{stance toward Democrats}, and \textit{stance toward Republicans}. The prompts used for annotation are shown in Figures~\ref{fig:prompt_ideo_bias} and \ref{fig:prompt_stance}.

To validate the LLM annotations, two annotators manually labeled a random sample of 25 news articles and their 75 summaries for all three measures of political bias. The annotators reached adequate inter-rater agreement on all three measures: \textit{ideological bias} (accuracy = 79\%, Krippendorff's $\alpha = .82$), \textit{stance toward Democrats} (accuracy = 93\%, $\alpha = .79$), and \textit{stance toward Republicans} (accuracy = 86\%, $\alpha = .77$). We evaluated the alignment between one annotator and the LLM using three metrics: (1)~accuracy, the raw percent agreement; (2)~$F_1$ score~\citep{derczynskiComplementarityFscoreNLP2016}, the harmonic mean of precision and recall, macro-averaged across classes; and (3) $O_1$ score, an ordinal extension of $F_1$ that penalizes misclassifications in proportion to their distance from the true label, as formulated in~\ref{app:sec:f1_o1}. The alignment between the annotator and the LLM is high for the three-point measures, \textit{stance toward Democrats} (accuracy = 92\%, $F_1$ = .87, $O_1$ = .93) and \textit{stance toward Republicans} (accuracy = 82\%, $F_1$ = .73, $O_1$ = .87); for the five-point measure of \textit{ideological bias}, the $F_1$ score is substantially lower, yet the accuracy and $O_1$ score are comparable (accuracy = 80\%, $F_1$ = .47, $O_1$ = .90), meaning that disagreements are concentrated among neighboring scale points.

\begin{figure}[h]
\centering
\fbox{
\begin{minipage}{0.8\linewidth}  
\small
*Task*

Your task is to classify the ideological bias expressed by the author of a news report. Ideology here is defined in the context of the United States political system. Specifically, determine whether the author’s framing, tone, or evaluative language reflects ideological bias: far left, left, center, right, or far right. Articles with no ideological bias are classified as center. Do not classify the ideology of people, groups, or viewpoints that the author merely quotes, paraphrases, or describes. Instead, focus on the perspective conveyed through the overall framing and presentation of the article.

*Reasoning*

Left- and right-wing reporting may reflect different stances on economic and social dimensions. Use the following rationales to determine the ideological leaning implied by the author’s framing or narrative.

1. Left and far-left reporting (typically from a liberal perspective):

- Economic dimension: Framing that favors economic redistribution, government spending on public welfare and social services, and policies promoting equality.

- Social dimension: Framing that supports immigration, reproductive rights, LGBTQ+ rights, and climate change mitigation.

Left and far-left reporting may support liberal perspectives and/or be negative toward conservative perspectives (e.g., criticize pro-life policies, free-market, and other right and far-right policies).

2. Right and far-right reporting (typically from a conservative perspective):

- Economic dimension: Framing that favors limited government intervention, free-market principles, lower taxes, and reduced public spending on social programs (often alongside support for increased military spending).

- Social dimension: Framing that emphasizes individual responsibility and the preservation of traditional social and cultural values.

Right- and far-right reporting may support conservative perspectives and/or be negative toward liberal perspectives (e.g., criticize pro-choice policies, high taxes, and other left and far-left policies).

Reports with no ideological bias are classified as center.

*Output format*

Your reply must be one of five numbers: -2 (= far left), -1 (= left of center), 0 (= center), 1 (= right of center), or 2 (= far right).

Clarifying degree of bias (-2 vs. -1, 1 vs. 2): 

- Moderate bias (-1 or 1): The article shows a noticeable preference for one side. Bias influences parts of the framing but is not overwhelming, and the language is not propagandistic or emotionally charged. May omit relevant alternative perspectives.

- Strong bias (-2 or 2): The article clearly and consistently favors one side. Language may be emotionally-charged or propagandistic. Countervailing facts or salient alternative perspectives are often omitted. Bias is pervasive and central to the framing.

*Input*

The report you have to classify: ``\{text\}''
\end{minipage}
}
\caption{Prompt for annotating \textit{ideological bias}.}
\label{fig:prompt_ideo_bias}
\end{figure}

\begin{figure}[h]
\centering
\fbox{
\begin{minipage}{0.8\linewidth}  
\small
*Task*

Your task is to classify whether a news report is positive, neutral, or negative toward [Democrats/Republicans]. You only need to evaluate the stance toward [Democrats/Republicans], not toward [Republicans/Democrats] or other political actors.

*Definition*

[Democrats/Republicans] include: 1) [Democratic/Republican] politicians, 2) the [Democratic/Republican] Party itself, and 3) policies, agendas, or actions proposed or supported by the [Democratic/Republican] Party.

*Reasoning*

- Positive: The report conveys direct or indirect praise or approval of [Democratic/Republican] politicians, agendas, policies, or actions, including explicit endorsements or subtle references to achievements, popularity, or successes. Reporting that suggests likely [Democratic/Republican] electoral victories or emphasizes favorable poll results in a positive framing counts as positive.

- Neutral: The report presents [Democrats/Republicans] in a factual or descriptive manner without clear approval or disapproval. The report may simply describe events, statements, or positions involving [Democrats/Republicans] without expressing a stance.

- Negative: The report conveys criticism or disapproval of the [Democratic/Republican] Party, its politicians, agendas, policies, or actions. This may include negative descriptions, highlighting failures or controversies in a critical way, or otherwise presenting [Democrats/Republicans] in an unfavorable light.

*Output format*

Your reply must be one of three numbers: -1 (= negative), 0 (= neutral), or 1 (= positive).

*Input*

The report you have to classify: ``\{text\}''
\end{minipage}
}
\caption{Prompt for annotating \textit{stance toward Democrats/Republicans}.}
\label{fig:prompt_stance}
\end{figure}

\paragraph{Negativity}
We annotated the negativity of each news article and summary using a transformer-based classifier~\cite{talaga2025negativity}. The classifier was based on the \texttt{XLM-RoBERTa-Large} model~\citep{conneauUnsupervisedCrosslingualRepresentation2020} and fine-tuned on over 6 million news posts published by news outlets on Facebook to detect negative sentiment in news. 
Specifically, text is labeled negative if the language conveys negative emotion (sadness, anger, fear, disgust), uses insults/incivility/sarcasm/derision or strongly critical evaluative wording, or contains rhetorical questions used to belittle or scorn. As with political bias, two annotators validated the negativity labels on the same sample of 25 articles and 75 summaries, achieving adequate inter-rater agreement (Krippendorf's $\alpha =.72$). One annotator's labels were then compared to the classifier output; the two showed strong alignment (accuracy = 85\%, $F_1=.85$). 

\paragraph{Negative emotions} We used GPT-5.2-2025-12-11 to annotate each news article and summary for \textit{anger} and \textit{fear}. The prompt used for annotation is shown in Figure~\ref{fig:prompt_emotion}. As with political bias, two annotators validated the LLM labels on the same sample of 25 articles and 75 summaries. The annotators reached adequate inter-rater agreement on both \textit{anger} (accuracy = 88\%, Krippendorff's $\alpha = .81$) and \textit{fear} (accuracy = 83\%, $\alpha = .71$). One annotator's labels were then compared to the LLM output; the two showed acceptable agreement for \textit{anger} (accuracy = 80\%, $F_1$ = .53, $O_1$ = .74) and \textit{fear} (accuracy = 74\%, $F_1$ = .75, $O_1$ = .88). 

\begin{figure}[h]
\centering
\fbox{
\begin{minipage}{0.8\linewidth}  
\small
Your task is to classify the level of [anger/fear] expressed in a news report: strong, moderate, or no [anger/fear] at all.

Definition of [anger/fear]: Expressions of [frustration, irritation, or rage/anxiety, apprehension, or dread].

Your reply must be one of three numbers: 0 (= no [anger/fear] at all), 1 (= moderate [anger/fear]), 2 (= strong [anger/fear]).

The report you have to classify: ``\{text\}''
\end{minipage}
}
\caption{Prompt for annotating \textit{anger/fear}.}
\label{fig:prompt_emotion}
\end{figure}

\paragraph{Journalistic writing quality}
We annotated the journalistic writing quality of each news article and summary using a transformer-based classifier. The classifier was developed to assess journalistic quality more holistically, covering targets such as topical importance, adequate sourcing, and writing quality (authors, unpublished). In this work, we employed the classifier on a subset of targets that we expected to vary meaningfully between articles and summaries. For example, we excluded dimensions pertaining to topical importance, as they reflect a property of the underlying event and should not differ between an article and its summary. We also excluded measures of adequate sourcing, as summaries necessarily score high on this dimension by attributing claims to the article they condense, rendering the scores uninformative. Instead, we focused on dimensions of \textit{journalistic writing quality}, which concern how ideas are expressed and how well core points are conveyed: clear writing and organization (\textit{clarity}), captivating and engaging writing (\textit{engagement}), avoidance of personal anecdotes in favor of more objective reporting (\textit{personal tone}), and the absence of clickbait and other sensationalistic or deceptive tactics (\textit{clickbait}).

To develop this classifier, two annotators initially coded 1{,}000 news articles by hand (Krippendorf's alpha between human annotators ranged from .75 to .95 for all dimensions except engagement, with $\alpha \approx .71$ for engagement)
The exact operational definitions of the above-mentioned target variables used by human coders are presented in Table~\ref{tab:quality_definitions}. \textit{Clarity} was coded as 0 (unclear or poorly structured) or 1 (clear and easy to follow), and \textit{engagement}, \textit{personal tone}, and \textit{clickbait} were coded as 0, 1 or 2, with 2 reflecting very vivid and engaging writing, high reliance on personal anecdotes or opinions, or heavy use of clickbait tactics. 

The 1{,}000 labeled articles were used to fine-tune an initial classifier based on \texttt{XLM-RoBERTa-Large} \citep{conneauUnsupervisedCrosslingualRepresentation2020}. As performance was not deemed adequate, this initial classifier was then used to purposively sample another 450 news articles, oversampling those with rare scores in each dimension, and stratifying by article source and length to ensure maximum variability within each rare-score group. One annotator labeled the additional 450 articles, after which the final classifier was produced by fine-tuning on the full set of 1{,}450 human-annotated examples. The final model achieved adequate accuracy across all four dimensions of \textit{clarity} (accuracy = 80\%, $F_1$ = .79), \textit{engagement} (accuracy = 75\%, $F_1$ = .58, $O_1$ = .79), \textit{personal tone} (accuracy = 71\%, $F_1$ = .69, $O_1$ = .83), and \textit{clickbait} (accuracy = 86\%, $F_1$ = .66, $O_1$ = .80).

To additionally ensure that this previously validated classifier performs adequately on the present data set, and in particular that it remains accurate when used to assess summaries instead of full articles, one researcher who initially annotated training data for the classifier manually validated its output for 25 news articles and their 75 summaries. As shown in Table~\ref{tab:quality_definitions}, the classifier was highly accurate on this sample (between 89\% and 99\%); and three of four measures received high $O_1$ scores ($F_1$ adjusted to account for the ordinal nature of the classifier). Clickbait, however, received a comparatively low $O_1$ score, despite only a single mismatch between human annotation and classifier. This result is driven by the incredible rarity of clickbait and sensationalism, which appear in only about 2\% of news articles and fewer than 0.6\% of summaries, making high $O_1$ scores difficult to achieve. 

\begin{table}[ht]
\centering
\small
\caption{Dimensions of journalistic writing quality, coding scheme, and classifier performance on the validation set of this work. Performance was measured based on three metrics: accuracy, the raw percent agreement; $F_1$ score~\citep{derczynskiComplementarityFscoreNLP2016}, macro-averaged across classes; and $O_1$ score, an ordinal extension of $F_1$, as formulated in~\ref{app:sec:f1_o1}.}
\renewcommand{\arraystretch}{1.3}
\begin{tabular}{|l|c|p{11cm}|l|}
\hline
Dimension & Score & Coding Scheme & Performance               \\ \hline
\multirow{6}{*}{Clarity}    & 0     & Not very clear. Some cases may include slang/emojis, asides that reduce clarity, awkward sequencing or jargon. More severe cases may be hard to follow or overloaded with unexplained jargon/technical terms such that an average adult reader would struggle to grasp the main ideas. Alternatively, it may be circuitous or bury important details late in the article.                                    & \multirow{2}{*}{\begin{tabular}[c]{@{}l@{}}Accuracy: 95\%\\ $F_1$: .76\\ $O_1$: .76 \end{tabular}} \\
                            & 1     & Clear and easy to follow. Main ideas are apparent even to skimmers or readers without topic expertise. Everyday vocabulary and coherent organization. The most important ideas are right at the beginning of the article.                   & \\ 
\hline
\multirow{6}{*}{Engagement} & 0     & The text is not engaging, it is boring even for most news readers, unless they already have a specific interest in the topic;           & \multirow{3}{*}{\begin{tabular}[c]{@{}l@{}}Accuracy: 89\%\\ $F_1$: .65\\ $O_1$ : .80\end{tabular}} \\
                            & 1     & The text is generally engaging for a person with interest in the broad topic or slightly engaging to most readers & \\
                            & 2     & The text is very engaging, it is vividly written and uses colorful language in ways that make it interesting to read even for a person who does not read news very often or who is not particularly interested in the topic.                        & \\ 
\hline
\multirow{6}{*}{Personal}   & 0     & The text is generally focused on reporting verifiable facts, does not contain personal opinions or anecdotes, and focuses on broad issues rather than individuals;                                      & \multirow{3}{*}{\begin{tabular}[c]{@{}l@{}}Accuracy: 90\%\\ $F_1$: .56\\ $O_1$ : .80\end{tabular}} \\
                            & 1     & The text is somewhat personal with some subjective opinions, but also contains verifiable facts and some objective reporting, or the text describes broader issues, but contextualizes them with human-interest elements such as individuals’ stories;             & \\
                            & 2     & The text contains primarily subjective opinions and perspectives or individual anecdotes and experiences.        & \\ 
\hline
\multirow{10}{*}{Clickbait}  & 0     & The text does not make use of unnecessary repetition, does not withhold core information until the end, does not use sensationalist framing or language, and the headline clearly and accurately represents the text’s contents;                               & \multirow{3}{*}{\begin{tabular}[c]{@{}l@{}}Accuracy: 99\%\\ $F_1$: .50\\ $O_1$ : .50\end{tabular}} \\
                            & 1     & The text shows some signs of clickbait or SEO focus, such as by using unnecessary repetition, exaggerated framing, or questions whose answers are delayed, but still provides some informative content. The headline may be framed as a question or use exaggerated language, but still allows readers to know what the main point will be;           & \\
                            & 2     & The text is strongly clickbait/SEO-focused. It may use a repetitive structure, delay answers to important questions, add irrelevant information, or include sensationalist writing; or it may have a headline that uses clickbait like a question, exaggeration, or forward-reference, or is otherwise misleading about the text’s contents. The text has little informational value compared to its length. & \\                                                  
\hline
\end{tabular}
\label{tab:quality_definitions}
\end{table}

\subsection{Topic Modeling}
\label{app:sec:topic}
We annotated the topic of each news article using a BERTopic model \cite{grootendorst2022bertopic}, in which articles are embedded with \texttt{all-MiniLM-L6-v2}, dimensionality is reduced via UMAP (\texttt{n\_neighbors=15, n\_components=5, random\_state=42}), clusters are identified via HDBSCAN (\texttt{min\_cluster\_size=20, cluster\_selection\_method=`leaf'}), and topic representations are derived using c-TF-IDF. The approach yielded 125 topic clusters, which we then manually aggregated into 19 broader topics, as shown in Table~\ref{tab:topics}.

\begin{table}[ht]
\centering
\small
\caption{Topics, clusters (represented by keywords), and article counts per topic.}
\renewcommand{\arraystretch}{1.3}
\begin{tabular}{|p{3.5cm}|p{11cm}|r|}
\hline
Topic & Clusters & \#Articles \\
\hline
Election & biden\_debate, harris\_kamala, term\_voters\_trumps, puerto rico\_comedy\_song, harris\_voters\_polls, secret service\_assassination, haley\_primary, voter\_ballots\_voting, walz\_minnesota, iowa\_desantis\_caucuses, mamdani\_cuomo\_mayor & 1763 \\
\hline
Trump Legal Cases & merchan\_trial\_sentencing, immunity\_smith\_special counsel, daniels\_cohen\_pecker, carroll\_kaplan\_defamation, willis\_wade\_giuliani & 434 \\
\hline
Government & musk\_tesla\_spacex, hegseth\_rubio\_chat, johnson\_speaker\_mcconnell, employees\_firings\_probationary, fbi\_patel\_intelligence, tax\_irs\_taxpayers, hur\_mayorkas\_impeachment, doge\_musk\_efficiency, pardon\_hunter, budget\_big beautiful, shutdown\_schumer\_cr, usaid\_foreign assistance, social security\_retirement, gaetz\_ethics committee, hunter biden & 1406 \\
\hline
Courts & justices\_courts\_supreme, chevron\_circuit\_loper, citizenship\_birthright\_injunctions & 314 \\
\hline
Immigration & immigration\_ice\_border, abrego garcia\_el salvador, texas\_abbott\_border, khalil\_ozturk\_detention & 528 \\
\hline
Israel--Mideast & gaza\_hamas\_hostages, iran\_nuclear, jews\_antisemitism, hezbollah\_lebanon, saudi\_qatar\_arabia, syria\_assad, houthis\_red sea\_yemen, genocide\_icj\_israel & 1234 \\
\hline
Russia--Ukraine & ukraine\_russia\_putin, navalny\_gershkovich\_assange & 557 \\
\hline
World Affairs & china\_beijing\_xi, afd\_macron\_european, reeves\_labour\_nhs, india\_pakistan\_modi, labour\_starmer\_sunak, canada\_trudeau, taiwan\_china\_philippines, police\_london\_southport, taliban\_afghanistan, maduro\_venezuela\_sheinbaum, m23\_rsf\_congo\_sudan, korea\_north korea\_kim, dutton\_australia\_albanese, south africa\_ramaphosa, greenland\_panama\_denmark, yoon\_martial law, rwanda\_asylum seekers, haiti\_gangs, myanmar\_earthquake, milei\_argentina & 1439 \\
\hline
Economy & tariffs\_trade\_imports, inflation\_fed\_rates, banks\_jpmorgan, sp 500\_stocks, jobs\_unemployment, oil\_barrels\_opec, egg\_prices\_food, bitcoin\_crypto\_sec, strike\_union\_transit, debt\_credit\_heloc, ev\_tesla\_electric, steel\_nippon, borrowers\_student loan & 1408 \\
\hline
Health & kennedy\_autism\_rfk, plastic\_fluoride\_pfas, medicaid\_medicare\_health care, cannabis\_marijuana\_overdose, covid\_vaccine\_fauci, measles\_flu\_bird flu, drugs\_medicare\_obesity, mpox\_malaria\_dengue & 659 \\
\hline
Environment & climate\_emissions\_epa, energy\_solar\_wind, fema\_hurricane\_helene, fires\_wildfires\_palisades & 415 \\
\hline
Crime & police\_shooting\_officers, combs\_diddy\_trafficking, adams\_mayor\_corruption, national guard\_newsom\_LA, mangione\_unitedhealthcare, jabbar\_orleans\_attack, menendez\_brothers & 772 \\
\hline
Education & campus\_students\_columbia, harvard\_international students, schools\_teachers\_education, education\_department\_mcmahon & 460 \\
\hline
Technology & ai\_google\_openai, npr\_fcc\_60 minutes, platforms\_social media, tiktok\_bytedance\_ban, cybersecurity\_hackers & 669 \\
\hline
Gender \& DEI & transgender\_genderaffirming, dei\_diversity\_pride, sports\_transgender athletes\_title ix & 310 \\
\hline
Abortion & abortion\_pregnancy\_ivf & 188 \\
\hline
Religion & book\_society\_atheism, pope\_francis\_vatican, church\_christian\_faith & 350 \\
\hline
Guns & firearms\_second amendment, school\_shooting\_columbine & 162 \\
\hline
Miscellaneous & housing\_rent\_homelessness, olympics\_ioc, season\_coach\_nfl, cemetery\_parade\_veterans, boeing\_airlines\_max, faa\_air traffic, leadership\_employees\_talent & 709 \\
\hline
\end{tabular}
\label{tab:topics}
\end{table}

\subsection{Data Descriptives}
\begin{figure}[ht]
\centering
\includegraphics[width=\linewidth]{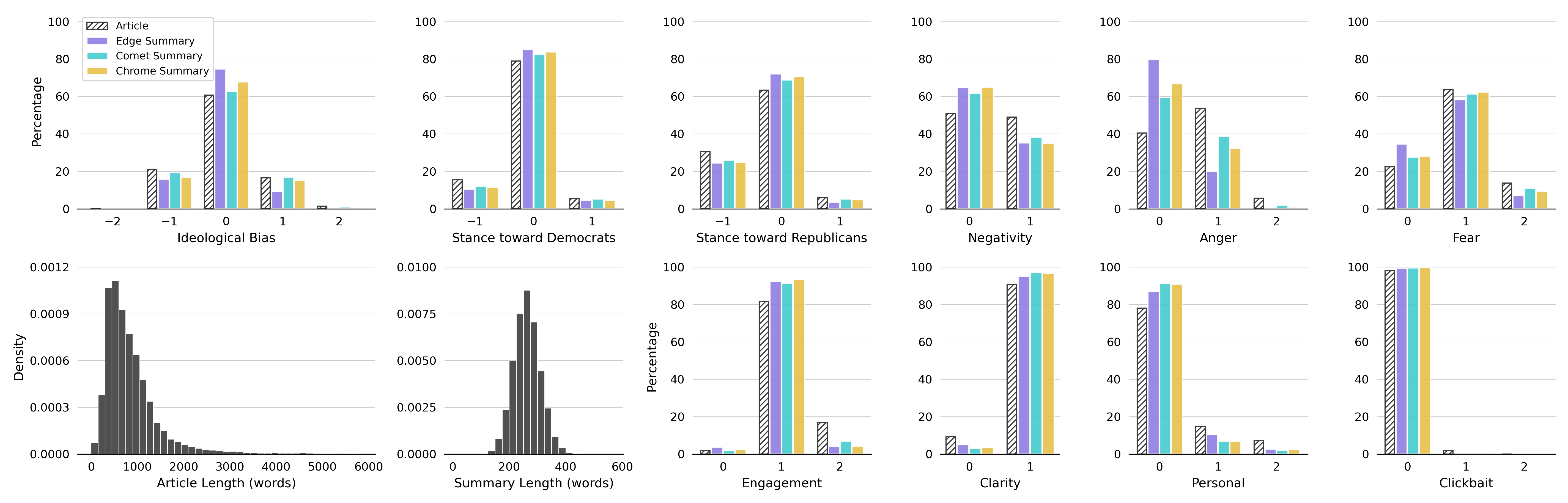}
\caption{Distributions of all variables, as well as article length and summary length (in words).}
\label{fig:dist}
\end{figure}

Fig.~\ref{fig:dist} plots the distributions of all annotated variables, as well as article and summary length. Table~\ref{tab:counts} reports the counts and percentages of articles, Edge summaries, Comet summaries, and Chrome summaries assigned each label for each variable.

\begin{table}[ht]
\centering
\small
\caption{Counts and percentages of articles, Edge summaries, Comet summaries, and Chrome summaries assigned each label for every variable ($N=13{,}777$).}
\sisetup{group-separator={,}, group-minimum-digits=4}
\newcolumntype{P}{S[table-format=2.1, table-space-text-post={\%}]}
\begin{tabular}{@{}l S[table-format=-1.0] S[table-format=5.0, table-column-width=1.2cm] P S[table-format=5.0, table-column-width=1.2cm] P S[table-format=5.0, table-column-width=1.2cm] P S[table-format=5.0, table-column-width=1.2cm] P@{}}
\toprule
 & & \multicolumn{2}{c}{Article} & \multicolumn{2}{c}{Edge} & \multicolumn{2}{c}{Comet} & \multicolumn{2}{c}{Chrome} \\
\cmidrule(lr){3-4}\cmidrule(lr){5-6}\cmidrule(lr){7-8}\cmidrule(lr){9-10}
Variable & {Label} & {Count} & {\%} & {Count} & {\%} & {Count} & {\%} & {Count} & {\%} \\
\midrule
\multirow{5}{*}{Ideological bias} & -2 & 15 & 0.1\% & 2 & 0.0\% & 0 & 0.0\% & 0 & 0.0\% \\
 & -1 & 2904 & 21.1\% & 2180 & 15.8\% & 2666 & 19.4\% & 2294 & 16.7\% \\
 & 0 & 8381 & 60.8\% & 10303 & 74.8\% & 8642 & 62.7\% & 9350 & 67.9\% \\
 & 1 & 2278 & 16.5\% & 1262 & 9.2\% & 2329 & 16.9\% & 2075 & 15.1\% \\
 & 2 & 199 & 1.4\% & 30 & 0.2\% & 140 & 1.0\% & 58 & 0.4\% \\
\midrule
\multirow{3}{*}{Stance toward Democrats} & -1 & 2136 & 15.5\% & 1437 & 10.4\% & 1664 & 12.1\% & 1599 & 11.6\% \\
 & 0 & 10889 & 79.0\% & 11710 & 85.0\% & 11399 & 82.7\% & 11554 & 83.9\% \\
 & 1 & 752 & 5.5\% & 630 & 4.6\% & 714 & 5.2\% & 624 & 4.5\% \\
\midrule
\multirow{3}{*}{Stance toward Republicans} & -1 & 4203 & 30.5\% & 3371 & 24.5\% & 3569 & 25.9\% & 3391 & 24.6\% \\
 & 0 & 8739 & 63.4\% & 9936 & 72.1\% & 9491 & 68.9\% & 9734 & 70.7\% \\
 & 1 & 835 & 6.1\% & 470 & 3.4\% & 717 & 5.2\% & 652 & 4.7\% \\
\midrule
\multirow{2}{*}{Negativity} & 0 & 7018 & 50.9\% & 8920 & 64.7\% & 8494 & 61.7\% & 8953 & 65.0\% \\
 & 1 & 6759 & 49.1\% & 4857 & 35.3\% & 5283 & 38.3\% & 4824 & 35.0\% \\
\midrule
\multirow{3}{*}{Anger} & 0 & 5579 & 40.5\% & 10989 & 79.8\% & 8185 & 59.4\% & 9195 & 66.7\% \\
 & 1 & 7407 & 53.8\% & 2751 & 20.0\% & 5343 & 38.8\% & 4464 & 32.4\% \\
 & 2 & 791 & 5.7\% & 37 & 0.3\% & 249 & 1.8\% & 118 & 0.9\% \\
\midrule
\multirow{3}{*}{Fear} & 0 & 3084 & 22.4\% & 4776 & 34.7\% & 3808 & 27.6\% & 3886 & 28.2\% \\
 & 1 & 8806 & 63.9\% & 8038 & 58.3\% & 8449 & 61.3\% & 8595 & 62.4\% \\
 & 2 & 1887 & 13.7\% & 963 & 7.0\% & 1520 & 11.0\% & 1296 & 9.4\% \\
\midrule
\multirow{2}{*}{Clarity} & 0 & 1284 & 9.3\% & 694 & 5.0\% & 406 & 2.9\% & 464 & 3.4\% \\
 & 1 & 12493 & 90.7\% & 13083 & 95.0\% & 13371 & 97.1\% & 13313 & 96.6\% \\
\midrule
\multirow{3}{*}{Engagement} & 0 & 242 & 1.8\% & 509 & 3.7\% & 238 & 1.7\% & 325 & 2.4\% \\
 & 1 & 11230 & 81.5\% & 12719 & 92.3\% & 12578 & 91.3\% & 12854 & 93.3\% \\
 & 2 & 2305 & 16.7\% & 549 & 4.0\% & 961 & 7.0\% & 598 & 4.3\% \\
\midrule
\multirow{3}{*}{Personal tone} & 0 & 10742 & 78.0\% & 11972 & 86.9\% & 12547 & 91.1\% & 12514 & 90.8\% \\
 & 1 & 2039 & 14.8\% & 1444 & 10.5\% & 960 & 7.0\% & 945 & 6.9\% \\
 & 2 & 996 & 7.2\% & 361 & 2.6\% & 270 & 2.0\% & 318 & 2.3\% \\
\midrule
\multirow{3}{*}{Clickbait} & 0 & 13509 & 98.1\% & 13693 & 99.4\% & 13713 & 99.5\% & 13717 & 99.6\% \\
 & 1 & 267 & 1.9\% & 84 & 0.6\% & 63 & 0.5\% & 60 & 0.4\% \\
 & 2 & 1 & 0.0\% & 0 & 0.0\% & 1 & 0.0\% & 0 & 0.0\% \\
\bottomrule
\end{tabular}
\label{tab:counts}
\end{table}

\subsection{$O_1$ score}\label{app:sec:f1_o1}

Here we define the $O_1$ measure for ordinal classification problems. Let us first define Relative Absolute Error (RAE) as:
\begin{equation}
    \text{RAE}(y, \hat{y}) = \frac{|y - \hat{y}|}{|\max\Omega_Y - \min\Omega_Y|}
\end{equation}
where $y$ and $\hat{y}$ are observed values of the target random variable $Y$ and predicted random variable $\hat{Y}$, and $\Omega_Y$ stands for the support (sample space) of $Y$. Now, we can define ordinally weighted precision and recall scores for label $l$ as appropriate conditional expectations for RAE:
\begin{alignat}{2}
    O\text{-Precision}(l) 
    &= \mathbb{E}_{\sim Y}\left[
        \text{RAE}(y, \hat{y}) \mid \hat{Y} = l
    \right] \label{app:eq:o-precision} \\
    O\text{-Recall}(l) 
    &= \mathbb{E}_{\sim\hat{Y}}\left[
        \text{REA}(y, \hat{y}) \mid Y = l
    \right]
\end{alignat}
 
Finally, we define the $O_1$ score for label $y$ as the harmonic mean of the ordinally-weighted precision and recall:
\begin{equation}\label{app:eq:o1}
    O_1(l) 
    = 2\frac{
        O\text{-Precision}(l)O\text{-Recall}(l)
    }{
        O\text{-Precision}(l) + O\text{-Recall}(l)
    }
\end{equation}
Overall $O_1$ score can be defined in several ways, but here we use \enquote{macro} averaging, so:
\begin{equation}\label{app:eq:o1-macro}
   O_1 = \frac{1}{|\Omega_Y|} \sum_{l \in \Omega_Y}O_1(l)
\end{equation}

\clearpage
\section{Additional Results}\label{app:sec:res}

\subsection{Accuracy}\label{app:sec:res-acc}
\begin{figure}[h]
\centering
\includegraphics[width=.55\linewidth]{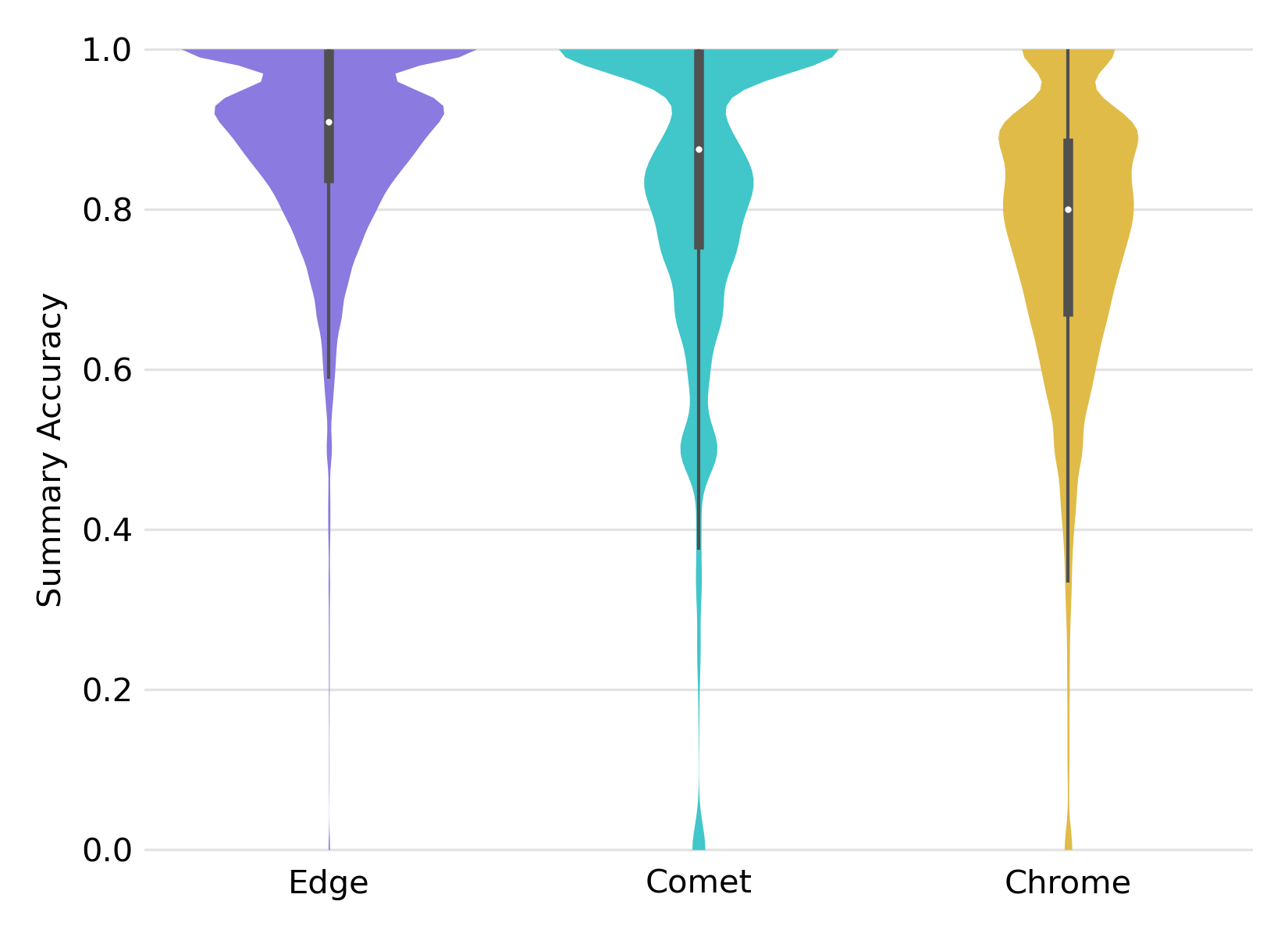}
\caption{Distribution of summary accuracy scores for each browser. Within each violin, the white dot denotes the median, the thick bar spans the interquartile range (Q1–Q3), and the thin line extends to 1.5 times the interquartile range.}
\label{fig:acc-dist}
\end{figure}

\paragraph{Accuracy distributions}
Fig.~\ref{fig:acc-dist} shows the distribution of summary accuracy scores for each browser. We see that most summaries are \textit{mostly accurate}, although not many are \textit{fully accurate} (25.3\% in Edge, 42.1\% in Comet, 12.1\% in Chrome).

\paragraph{Accuracy across outlets}
Table~\ref{tab:acc-outlet} reports the average summary accuracy (in percentage) for each news outlet and browser. We see that articles from Reason, MSNBC, and Forbes consistently receive more accurate summaries, while those from Fox News and The Guardian consistently receive less accurate ones.

\begin{table}[h]
\centering
\small
\caption{Average summary accuracy by news outlet and browser, sorted from highest to lowest by mean accuracy across browsers.}
\begin{tabular}{lrrrr}
\toprule
Outlet & Edge (\%) & Comet (\%) & Chrome (\%) & Mean (\%) \\
\midrule
Reason                  & 90.5 & 86.9 & 83.5 & 87.0 \\
MSNBC                   & 90.2 & 85.9 & 81.3 & 85.8 \\
Forbes                  & 89.9 & 87.6 & 79.4 & 85.6 \\
The Free Press          & 87.6 & 85.7 & 79.4 & 84.2 \\
The New York Times      & 89.0 & 83.5 & 79.0 & 83.8 \\
The Wall Street Journal & 88.4 & 86.1 & 75.7 & 83.4 \\
Reuters                 & 88.9 & 83.4 & 77.9 & 83.4 \\
CNN                     & 88.3 & 83.0 & 78.2 & 83.2 \\
The Hill                & 87.6 & 83.6 & 77.0 & 82.7 \\
NBC News                & 88.6 & 81.2 & 76.4 & 82.1 \\
Associated Press        & 87.5 & 83.1 & 74.7 & 81.8 \\
CBS News                & 88.5 & 82.6 & 71.4 & 80.8 \\
OANN                    & 87.6 & 78.9 & 72.8 & 79.7 \\
The Guardian            & 86.5 & 81.0 & 71.4 & 79.6 \\
Fox News                & 86.4 & 78.5 & 72.0 & 79.0 \\
\bottomrule
\end{tabular}
\label{tab:acc-outlet}
\end{table}

\paragraph{Statistical test of accuracy.}
To test for the effect of browser, outlet leaning, and topic on the accuracy of AI news summaries while accounting for summary length and article-level variation, we fit the following mixed-effects ordered beta regression model \cite{kubinec2023ordered} using the \texttt{glmmTMB} package \cite{brooks2017glmmtmb} in \texttt{R}:
$$\text{accuracy} \sim \text{browser} + \text{outlet\_lean} + \text{topic} + \text{log(article\_length)} + \text{summary\_length} + (1 \mid \text{outlet}) + (1 \mid \text{article\_id})$$
Accuracy ratio is modeled with an ordered beta family to accommodate a response bounded on [0, 1] with mass at the endpoints. Browser (reference: \textit{Chrome}), outlet leaning (reference: \textit{Center}), and topic (reference: \textit{Miscellaneous}) are included as categorical fixed effects. Standardized article length (log-scaled) and summary length are included as controls. Random intercepts for outlet and article account for non-independence among summaries for the same outlet and the same source article, with articles nested within outlets. The results are shown in Table~\ref{tab:reg-acc}. 

\begin{table}[t]
\centering
\small
\caption{Mixed-effects ordered beta regression of summary accuracy. Estimates on the logit scale. Reference categories: Chrome (browser), Center (outlet leaning), Miscellaneous (topic). Random intercepts: outlet, article. $N=41{,}293$.
$^{*}p<.05$, $^{**}p<.01$, $^{***}p<.001$.}
\label{tab:reg-acc}
\begin{tabular}{l S[table-format=-1.3, table-column-width=1.1cm] @{}l S[table-format=1.3, table-column-width=1.1cm]}
\toprule
{Predictor} & \multicolumn{2}{c}{Estimate} & {SE} \\
\midrule
Intercept & 1.081 & $^{***}$ & 0.043 \\
\addlinespace
\multicolumn{4}{l}{\textit{Browser}} \\
\quad Edge  & 0.441 & $^{***}$ & 0.010 \\
\quad Comet & 0.245 & $^{***}$ & 0.010 \\
\addlinespace
\multicolumn{4}{l}{\textit{Outlet Leaning}} \\
\quad Left  & -0.031 & & 0.055 \\
\quad Right & -0.010 & & 0.055 \\
\addlinespace
\multicolumn{4}{l}{\textit{Topic}} \\
\quad Abortion          & -0.063 &                       & 0.040 \\
\quad Courts            & 0.055 &                       & 0.035 \\
\quad Crime             & -0.022 &                       & 0.026 \\
\quad Economy           & 0.048 & $^{*}$   & 0.023 \\
\quad Education         & -0.062 & $^{*}$   & 0.029 \\
\quad Election          & -0.008 &                       & 0.022 \\
\quad Environment       & -0.071 & $^{*}$   & 0.030 \\
\quad Gender \& DEI     & 0.008 &                       & 0.034 \\
\quad Government        & 0.008 &                       & 0.023 \\
\quad Guns              & -0.087 & $^{*}$   & 0.043 \\
\quad Health            & -0.010 &                       & 0.027 \\
\quad Immigration       & -0.032 &                       & 0.028 \\
\quad Israel-Mideast    & -0.019 &                       & 0.023 \\
\quad Religion          & 0.018 &                       & 0.033 \\
\quad Russia-Ukraine    & 0.003 &                       & 0.028 \\
\quad Technology        & 0.003 &                       & 0.027 \\
\quad Trump Legal Cases & 0.019 &                       & 0.030 \\
\quad World Affairs     & -0.053 & $^{*}$   & 0.023 \\
\addlinespace
\multicolumn{4}{l}{\textit{Control}} \\
\quad Article length    & 0.056 & $^{***}$ & 0.005 \\
\quad Summary length    & 0.160 & $^{***}$ & 0.006 \\
\addlinespace
\midrule
\addlinespace
\textit{Random effects} & \multicolumn{3}{c}{SD} \\
\quad Outlet  & \multicolumn{3}{c}{0.086} \\
\quad Article & \multicolumn{3}{c}{0.274} \\
\bottomrule
\end{tabular}
\end{table}

\paragraph{Qualitative analysis of inaccuracies}
On the same sample of 880 sentences from 90 summaries that we used to validate the accuracy labels, we conduct a qualitative analysis to identify types of inaccuracies leading to \textit{unverifiable} or \textit{contradicted} labels, as well as their frequencies. The annotator identifies 19 contradicted sentences (2.2\%) and 144 unverifiable sentences (16.4\%) in the sample, rates roughly aligning with the 2.5\% contradicted and 14.7\% unverifiable rates across the entire LLM-annotated dataset. Table~\ref{tab:inacc} presents the identified types of inaccuracies, their frequencies, and illustrative examples; the types are ordered by increasing severity.

We see that \textit{made-up claims}, the most serious inaccuracies, occur extremely rarely in both contradicted and unverifiable sentences: only 1 contradicted sentence (5.3\%) and 8 unverified sentences (5.6\%) contain entirely fabricated, baseless claims that are neither stated nor implied by the source; these 9 sentences are spread across 9 distinct summaries, leading to 9 out of 90 (10\%) summaries containing at least one made-up claim. The largest share of inaccuracies in both contradicted (57.9\%) and unverifiable sentences (27.8\%) instead arises from \textit{attribution errors}, which are often introduced when AI summarizers attempt to bundle claims about distinct subjects: for example, ``A said X'' and ``B said Y'' may be summarized as ``A and B said X and Y'', with actions misattributed across subjects. In addition, \textit{inaccurate wording} accounts for 31.6\% of contradicted facts and 14.6\% of unverifiable ones. Unverifiable facts also stem from overclaimed or overinterpreted implied meaning (25.7\%), contextual information added by the summarizer (13.2\%), or summarization of other content on the webpage beyond the article text (13.2\%). Notably, as shown in Table~\ref{tab:inacc}, one contradicted sentence contains a fact that is incorrectly stated in the article and corrected by the summary.

Based on the qualitative analysis, roughly 95\% of inaccuracies in AI news summaries reflect imprecise phrasing in summarization, overreaching interpretations of source facts, or additional context that is largely accurate but beyond the scope of the source. Only around 5\% of inaccurate sentences---or an estimated 0.9\% of all sentences in AI news summaries---contain entirely fabricated facts.

\begin{table}[h]
\centering
\small
\caption{Types of inaccuracies in contradicted and unverifiable sentences from AI news summaries, their frequencies of occurrence, and illustrative examples. Types are ordered by increasing severity.}
\begin{tabular}{lp{9.2cm}rrr}
\toprule
Type & Example & Count & \% & Accum. \% \\
\midrule
\multicolumn{5}{l}{\textit{Contradicted}} \\
\quad Corrected fact    & \textcolor{blue!90}{\textit{Fauci is set to release \textbf{a book titled ``On Call: A Doctor's Journey in Public Service''} (the article references an excerpt titled ``He Loves Me, He Loves Me Not'' ...}} & 1 & 5.3\% & 5.3\%  \\
& The article claims that \textit{Fauci has a book entitled ``He Loves Me, He Loves Me Not''}, which is actually the name of a chapter in his book, and the correct book title is ``On Call: A Doctor's Journey in Public Service''. &&& \\
\addlinespace
\quad Attribution error     & \textcolor{blue!90}{\textit{A book \textbf{by the authors} was removed from ...}} & 11 & 57.9\% & 63.2\%  \\
& The book was written by one of the article authors and their other co-authors, not by authors of the article. &&&\\
\addlinespace
\quad Inaccurate wording & \textcolor{blue!90}{\textit{Trump \textbf{has not attempted} to reconcile with Haley or her supporters.}} & 6 & 31.6\% & 94.7\%  \\
& The source article states ``has made few attempts'', not none. &&& \\
\addlinespace
\quad Made-up claim       & \textcolor{blue!90}{\textit{Coverage includes Trump’s ongoing legal troubles, including his Manhattan criminal trial over alleged hush-money payments in 2016 \textbf{and separate Michigan proceedings involving ``fake electors'' who backed him in 2020}.}} & 1 & 5.3\% & 100.0\% \\
& The source article discusses Trump’s Manhattan criminal trial but does not mention any Michigan ``fake electors'' proceedings. &&& \\
\addlinespace
\multicolumn{2}{l}{\textbf{\quad Subtotal}} & \textbf{19} & & \\
\midrule
\multicolumn{5}{l}{\textit{Unverifiable}} \\
\quad Other info on the webpage                              & \textcolor{blue!90}{\textit{A Monday morning fire in San Mateo County destroyed a major affordable housing project in \textbf{Redwood City}, ...}} & 19 & 13.2\% & 13.2\%  \\
& The location ``Redwood City'' is mentioned in the title of an embedded video on the webpage, but not in the article text. &&& \\
\addlinespace
\quad Added context (accurate)                               & \textcolor{blue!90}{\textit{\textbf{Indian benchmarks} Sensex and Nifty, after hitting record highs until September last year, ...}}& 19 & 13.2\% & 26.4\%  \\
& The source article mentions Sensex and Nifty without noting that they are Indian, which they indeed are. &&& \\
\addlinespace
\quad Attribution error                          & \textcolor{blue!90}{\textit{\textbf{Rights advocates and local officials} had anticipated that around 200 ICE agents would start pre-dawn sweeps in Chicago targeting people heading to work, with similar immigration enforcement planned in New York and Miami and potentially lasting several days.}} & 40 & 27.8\% & 54.2\%  \\
& In the source article, details including ``around 200 ICE agents'' and ``pre-dawn sweeps'' are attributed to one specific rights advocate, and only the general expectation of multi-city sweeps is attributed to local officials. &&& \\
\addlinespace
\quad Overclaim/Overinterpretation            & \textcolor{blue!90}{\textit{The list of removed titles appears to be based on keywords \textbf{related to gender and diversity rather than academic merit}.}} & 37 & 25.7\% & 79.9\%  \\
& The source article says the removed books appears to be identified based on keywords, but the ``related to gender and diversity'' and ``rather than academic merit'' parts are rather implied by the context. &&& \\
\addlinespace
\quad Inaccurate wording & \textcolor{blue!90}{\textit{Russian disinformation efforts appear more effective in 2024, especially in nations with large \textbf{Russian-speaking} minorities.}} & 21 & 14.6\% & 94.4\%  \\
& The article says ``Russian minorities'' instead of ``Russian-speaking minorities''. &&& \\
\addlinespace
\quad Made-up claim & \textcolor{blue!90}{\textit{Trump declined to provide a specific start date or confirm the targeted cities \textbf{due to security concerns}.}} & 8 & 5.6\% & 100.0\% \\
& The source article does not mention or imply any reason for Trump's refusal to name the start date or target cities. &&& \\
\addlinespace
\multicolumn{2}{l}{\textbf{\quad Subtotal}} & \textbf{144} & & \\
\bottomrule
\end{tabular}
\label{tab:inacc}
\end{table}

\subsection{Political Bias}\label{app:sec:res-ideo}
\begin{figure}[h]
\centering
\includegraphics[width=.95\linewidth]{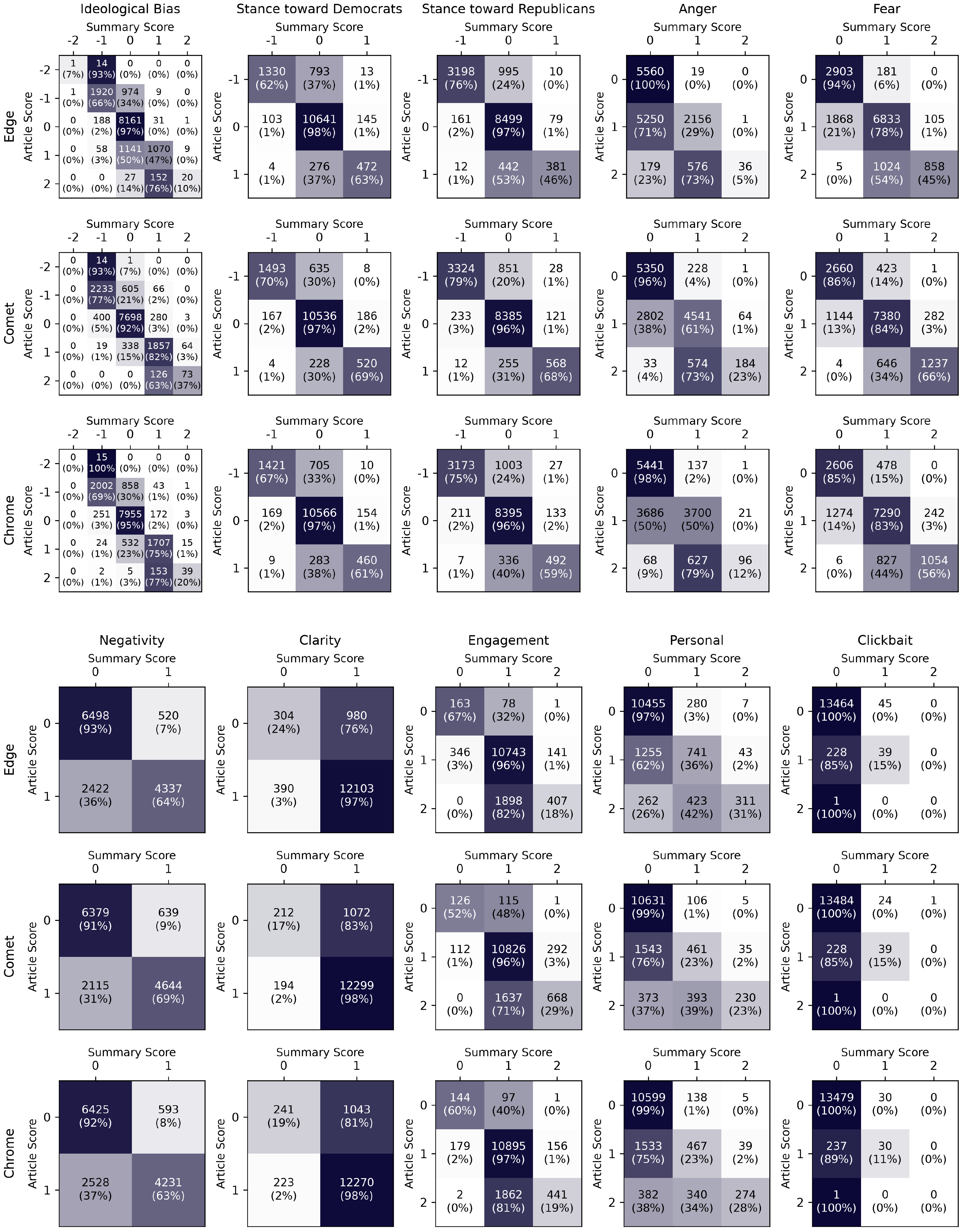}
\caption{Counts and percentages of article-to-summary score transitions for each browser and variable.}
\label{fig:transition}
\end{figure}

\paragraph{Score transitions in political bias}
Counts of article-to-summary score transitions for each browser and measure are visualized in Fig.~\ref{fig:transition}.

\paragraph{Statistical test of political bias reduction}
We test for the effect of browser, article leaning, and topic on political bias reduction by AI summarizers. For each measure of political bias (i.e., \textit{ideological bias}, \textit{stance toward Democrats}, and \textit{stance toward Republicans}), we fit the following mixed-effects logistic regression model using the \texttt{glmmTMB} package in \texttt{R}:
$$\text{reduced} \sim \text{browser} \times \text{article\_lean} + \text{outlet\_lean} + \text{cross\_cut} + \text{topic} + \text{log(article\_length)} + (1 \mid \text{outlet}) + (1 \mid \text{article\_id})$$

For each measure of political bias, the model is fit to summaries of articles with a non-neutral bias, and \textit{reduced} is a binary outcome indicating whether the summary exhibits reduced bias relative to its source article. Browser (reference: \textit{Chrome}), article leaning (reference: \textit{Left}), outlet leaning (reference: \textit{Center}), and topic (reference: \textit{Miscellaneous}) are included as categorical fixed effects, with an interaction between browser and article leaning. \textit{cross\_cut} is a binary predictor indicating whether the article's leaning runs counter to its outlet's leaning. Standardized article length (log-scaled) is included as a control. Random intercepts for outlet and article account for non-independence among summaries for the same outlet and the same source article, with articles nested within outlets. The results are shown in Table~\ref{tab:reg-ideo}. 

\begin{table}[t]
\centering
\small
\caption{Mixed-effects logistic regression of reduced political bias, fit separately for each measure. Estimates on the logit scale. Reference categories: Chrome (browser), Left (article leaning), Center (outlet leaning), Miscellaneous (topic). Random intercepts: outlet, article. Ideological bias: $N=16{,}188$, stance toward Democrats: $N=8{,}664$, stance toward Republicans: $N=15{,}114$.
$^{*}p<.05$, $^{**}p<.01$, $^{***}p<.001$.}
\label{tab:reg-ideo}
\begin{tabular}{l @{\hspace{25pt}} S[table-format=-1.3,table-column-width=1.1cm] @{}l S[table-format=1.3,table-column-width=1.1cm] S[table-format=-1.3,table-column-width=1.1cm] @{}l S[table-format=1.3,table-column-width=1.1cm] S[table-format=-1.3,table-column-width=1.1cm] @{}l S[table-format=1.3,table-column-width=1.1cm]}
\toprule
 & \multicolumn{3}{c}{Ideological bias} & \multicolumn{3}{c}{Stance toward Democrats} & \multicolumn{3}{c}{Stance toward Republicans} \\
\cmidrule(lr){2-4} \cmidrule(lr){5-7} \cmidrule(lr){8-10}
{Predictor} & \multicolumn{2}{c}{Estimate} & {SE} & \multicolumn{2}{c}{Estimate} & {SE} & \multicolumn{2}{c}{Estimate} & {SE} \\
\midrule
Intercept & -0.238 & & 0.463 & -0.754 &   & 0.408 & -1.600 & $^{***}$ & 0.405 \\
\addlinespace
\multicolumn{10}{l}{\textit{Browser}} \\
\quad Comet & -0.825 & $^{***}$ & 0.082 & -0.643 & $^{***}$ & 0.154 & -0.404 & $^{***}$ & 0.073 \\
\quad Edge  &  0.338 & $^{***}$ & 0.077 & -0.079 &                       & 0.150 & -0.020 &                       & 0.071 \\
\addlinespace
\multicolumn{10}{l}{\textit{Article leaning}} \\
\quad Right & -0.189 & & 0.211 & -0.468 & $^{*}$ & 0.215 & 1.310 & $^{***}$ & 0.222 \\
\addlinespace
\multicolumn{10}{l}{\textit{Outlet leaning}} \\
\quad Left  & -1.453 & $^{*}$  & 0.570 &  0.268 &                               & 0.218 & -0.441 & & 0.314 \\
\quad Right & -1.727 & $^{**}$ & 0.596 & -0.320 &   & 0.178 & -0.118 & & 0.366 \\
\addlinespace
\multicolumn{10}{l}{\textit{Interaction}} \\
\quad Comet $\times$ Right & -0.051 &                       & 0.121 & 0.333 &   & 0.181 & -0.386 & $^{*}$   & 0.159 \\
\quad Edge $\times$ Right  &  1.601 & $^{***}$ & 0.115 & 0.447 & $^{*}$           & 0.176 &  0.992 & $^{***}$ & 0.155 \\
\quad Cross-cut & 2.188 & $^{***}$ & 0.242 & 0.142 & & 0.235 & 0.728 & $^{**}$ & 0.229 \\
\addlinespace
\multicolumn{10}{l}{\textit{Topic}} \\
\quad Abortion          & -0.137 &                       & 0.333 & -0.160 &                       & 0.568 & -1.254 & $^{**}$  & 0.469 \\
\quad Courts            &  1.017 & $^{***}$ & 0.291 &  1.439 & $^{**}$  & 0.557 & -0.334 &                       & 0.430 \\
\quad Crime             &  0.380 &                       & 0.299 & -0.617 &                       & 0.492 & -0.254 &                       & 0.463 \\
\quad Economy           &  0.230 &                       & 0.254 &  0.017 &                       & 0.443 & -0.998 & $^{**}$  & 0.377 \\
\quad Education         & -0.330 &                       & 0.266 &  1.401 & $^{**}$  & 0.515 & -0.261 &                       & 0.432 \\
\quad Election          &  0.361 &                       & 0.225 & -1.388 & $^{***}$ & 0.384 & -0.387 &                       & 0.349 \\
\quad Environment       & -0.775 & $^{**}$  & 0.295 & -0.019 &                       & 0.493 & -0.257 &                       & 0.443 \\
\quad Gender \& DEI     & -0.951 & $^{***}$ & 0.281 &  0.283 &                       & 0.504 & -0.010 &                       & 0.432 \\
\quad Government        &  0.186 &                       & 0.231 & -0.349 &                       & 0.401 & -0.748 & $^{*}$   & 0.355 \\
\quad Guns              &  0.056 &                       & 0.389 &  0.386 &                       & 0.646 &  0.884 &                       & 0.587 \\
\quad Health            & -0.001 &                       & 0.266 &  0.106 &                       & 0.462 &  0.174 &                       & 0.403 \\
\quad Immigration       & -0.737 & $^{**}$  & 0.263 &  0.218 &                       & 0.466 & -0.828 & $^{*}$   & 0.384 \\
\quad Israel-Mideast    &  0.577 & $^{*}$   & 0.242 &  0.154 &                       & 0.443 &  0.407 &                       & 0.405 \\
\quad Religion          &  0.018 &                       & 0.286 &  0.501 &                       & 0.514 & -0.391 &                       & 0.454 \\
\quad Russia-Ukraine    &  0.974 & $^{**}$  & 0.319 &  0.743 &                       & 0.564 &  0.848 & $^{*}$   & 0.417 \\
\quad Technology        & -0.035 &                       & 0.280 &  0.584 &                       & 0.500 & -0.118 &                       & 0.417 \\
\quad Trump Legal Cases &  0.788 & $^{**}$  & 0.289 &  0.253 &                       & 0.537 & -1.938 & $^{***}$ & 0.394 \\
\quad World Affairs     &  0.634 & $^{*}$   & 0.254 &  1.459 & $^{**}$  & 0.479 &  1.107 & $^{**}$  & 0.385 \\
\addlinespace
\multicolumn{10}{l}{\textit{Control}} \\
\quad Article length & -0.059 & & 0.042 & 0.346 & $^{***}$ & 0.066 & 0.270 & $^{***}$ & 0.051 \\
\addlinespace
\midrule
\multicolumn{1}{l}{\textit{Random effects}} & \multicolumn{3}{c}{SD} & \multicolumn{3}{c}{SD} & \multicolumn{3}{c}{SD} \\
\quad Article & \multicolumn{3}{c}{1.989} & \multicolumn{3}{c}{2.589} & \multicolumn{3}{c}{2.339} \\
\quad Outlet  & \multicolumn{3}{c}{0.858} & \multicolumn{3}{c}{0.000} & \multicolumn{3}{c}{0.455} \\
\bottomrule
\end{tabular}
\end{table}

\paragraph{Qualitative analysis of asymmetry in political bias reduction}
To investigate whether the asymmetry in partisan stance attenuation reflects genuine political bias in AI summarization tools, we qualitatively analyze a random sample of 20 articles and their 60 summaries (5 articles and 15 summaries per category: pro-Republican, anti-Republican, pro-Democratic, anti-Democratic). The analysis reveals two recurring patterns: first, compared to anti articles, pro articles are generally easier to neutralize through summarization. Pro stances are often conveyed through positive framing, selective emphasis, or one-sided quotation, which can be easily adjusted for in summaries; by contrast, anti stances are often embedded in substantive criticism of politicians, parties, or policies, and therefore tend to be preserved in summaries. Second, compared to anti-Democratic articles, anti-Republican articles are less likely to receive neutral summaries. Anti-Republican articles tend to center entirely on criticism of Republicans, whereas anti-Democratic themes more often appear as secondary elements within pro-Republican articles, making them easier to downweight in summaries. In line with this observation, 27\% of anti-Democratic articles are also labeled pro-Republican, while only 14\% of anti-Republican articles are labeled pro-Democratic. Therefore, the observed asymmetry in partisan stance attenuation at least partially stems from pre-existing asymmetries in how partisan stances are expressed in the source articles. 

\paragraph{Qualitative analysis of Edge's political bias}
We investigate why Edge attenuates pro-Republican stances substantially more aggressively than the other two browsers. A reading of 15 summary sets---each set containing three summaries generated for the same article---reveals cases where Edge adopts a notably different summarization style for pro-Republican content. For example, where Comet summarizes an article as \textit{argues that blaming Trump for xx is misleading}, Edge renders the same article as \textit{discusses whether Trump is responsible for xxx and argues that xx’s criticism oversimplifies key facts}, which is phrased in a markedly more neutral tone. More tellingly, while Edge typically opens summaries with \textit{Here is a clear, concise summary of ...}, in 3.5\% of all cases it replaces ``concise'' with ``neutral''; however, this ``neutral'' variant appears in 6.6\% of Edge summaries for pro-Republican articles, but only in 1.9\% of summaries for pro-Democratic articles. Together, these patterns suggest that Edge may indeed exhibit a bias toward imposing stronger attenuation on pro-Republican content. 

\paragraph{Change in political bias across outlet leanings}
As shown in Fig.~\ref{fig:main-ideo}B, the magnitude of political bias reduction varies across outlet leanings. The most pronounced pattern emerges for cross-cutting articles, i.e., those whose position runs counter to their outlet's leaning. For ideological bias, 45\% of left-leaning articles published in right-leaning outlets receive more neutral summaries, versus 41\% in centrist outlets. Similarly, right-leaning articles in left-leaning outlets are more likely to receive more neutral summaries than right-leaning articles in centrist outlets (62\% vs. 42\%). Despite the limited number of cross-cutting articles, this effect is statistically significant (p<.001).

This pattern also emerges for cross-cutting partisan stances: pro-Democratic articles in right-leaning outlets are more likely to receive more neutral summaries (right: 37\% vs. centrist: 27\%), and anti-Democratic articles in left-leaning outlets are more likely to receive more neutral summaries (left: 40\% vs. centrist: 34\%). The same holds for pro-Republican articles in left-leaning outlets (left: 50\% vs. centrist: 39\%) and anti-Republican articles in right-leaning outlets (right: 31\% vs. centrist: 23\%). This cross-cutting effect is statistically significant for stance toward Republicans (p<.01). 

We also see that AI news summarization tools are more likely to shift neutral articles toward their outlet's leaning than away from it. This tendency is consistent across all three measures and strongest for ideological bias: 6\% of neutral articles from left-leaning outlets receive left-leaning summaries, while almost none receive right-leaning summaries; 9\% of neutral articles from right-leaning outlets receive right-leaning summaries, while only 1\% receive left-leaning summaries. These patterns point to the possibility that AI summarizers can be influenced by outlet leaning, although they may also stem from genuine differences in article content across outlets.

\begin{figure*}[htbp]
\centering
\includegraphics[width=.85\linewidth]{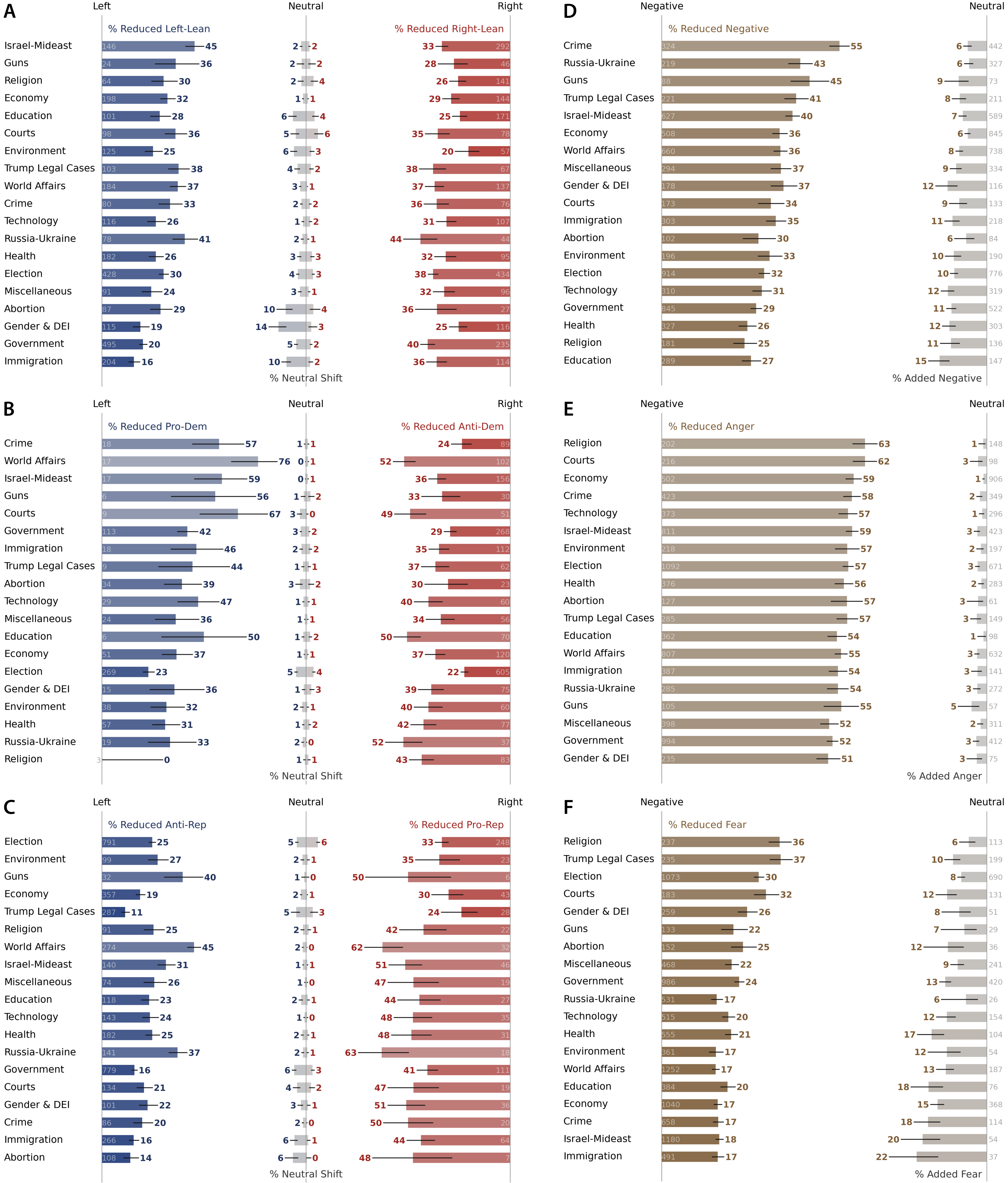}
\caption{(A-C) Change in political bias by topic, as measured by (A) \textit{ideological bias}, (B) \textit{stance toward Democrats}, and (C) \textit{stance toward Republicans}. In each panel, topics are ordered by the \textit{partisan shift} they induce, from most right-shifting (top) to most left-shifting (bottom). The partisan shift is calculated as $(p_{L\to N}-p_{R\to N})+(p_{N\to R}-p_{N\to L})$, where $p_{X\to Y}$ is the share of articles leaning $X$ whose summaries lean $Y$, and $L/N/R$ stands for left/neutral/right. Intuitively, a topic is more right-shifting if more left-leaning articles than right-leaning ones are pulled toward neutral, and more neutral articles are pulled toward the right than the left. Blue bars extending from the left show the share of left-leaning/pro-Democratic/anti-Republican articles receiving more neutral summaries, and red bars extending from the right show the share of right-leaning/anti-Democratic/pro-Republican articles receiving more neutral summaries. Gray bars in the middle show the share of neutral articles receiving left-leaning (toward the left) or right-leaning (toward the right) summaries. (D-F) Change in negative affect by topic, as measured by (D) \textit{negativity}, (E) \textit{anger}, and (F) \textit{fear}. In each panel, topics are ordered by the \textit{neutrality shift} they induce, from most neutral-shifting (top) to least (bottom). The neutrality shift is calculated as the share of negative articles receiving more neutral summaries minus the share of neutral articles receiving more negative summaries. Bars extending from the left show the share of articles with negativity/anger/fear receiving more neutral summaries, and bars extending from the right show the share of articles without negativity/anger/fear receiving summaries with more negative affect. Small numbers within/beside bars are article counts. Error bars indicate 95\% confidence intervals.}
\label{fig:topic}
\end{figure*}

\paragraph{Change in political bias across topics}
Fig.~\ref{fig:topic}A illustrates the change in political bias across topics. Two notable patterns emerge. First, news about international affairs are among the most likely to receive summaries of reduced political bias. Relative to the Miscellaneous baseline, this effect is statistically significant for World Affairs (\textit{ideological bias}, p<.05; \textit{stance toward Democrats}, p<.01; \textit{stance toward Republicans}, p<.01), Russia-Ukraine (\textit{ideological bias}, p<.01; \textit{stance toward Democrats}, p<.05), as well as Israel-Mideast (\textit{ideological bias}, p<.05). Second, a topic's \textit{partisan shift} in AI summaries---how far summaries move rightward or leftward relative to the source articles (see formulation in Fig.~\ref{fig:topic})---appears to align with \textit{issue ownership}, the perception that a given party is more \textit{competent} at handling certain issues over the long term~\citep{petrocik_issue_1996}, or, alternatively, that a given party is more \textit{associated} with specific issues~\citep{walgrave_associative_2012}. Specifically, for a set of topics typically considered Democratic-owned, including Gender \& DEI, and Abortion~\citep{wright_limits_2022, yang_striking_2026}, AI summaries shift left: ideologically neutral articles receive left-leaning summaries more often than right-leaning ones (left/right: 14\%/3\% for Gender \& DEI, 10\%/4\% for Abortion), and right-leaning articles are more often shifted toward neutral than left-leaning ones (right/left: 25\%/19\% for Gender \& DEI, 36\%/29\% for Abortion). By contrast, for a set of Republican-owned topics including war and national security (in our case, Israel-Mideast), Guns, Religion, and the Economy~\citep{wright_limits_2022, yang_striking_2026}, AI summaries are among the most right-shifting, with left-leaning articles more often shifted toward neutral than right-leaning ones (left/right: 45\%/33\% for Israel-Mideast, 36\%/28\% for Guns, 30\%/26\% for Religion, 32\%/29\% for Economy). That said, Immigration---traditionally considered to be Republican owned---is among the most left-shifted topics (10\%/2\% of ideologically neutral articles shift left/right, and 16\%/36\% left-/right-leaning articles shift neutral).

\subsection{Negative Affect}\label{app:sec:res-emo}
\paragraph{Score transitions in negative affect}
Counts of article-to-summary score transitions for each browser and measure are visualized in Fig.~\ref{fig:transition}.

\paragraph{Statistical test of negative affect reduction}
We test for the effect of browser, article leaning, and topic on negative affect reduction by AI summarizers. For each measure of negative affect (i.e., \textit{negativity}, \textit{anger}, and \textit{fear}), we fit the following mixed-effects logistic regression model using the \texttt{glmmTMB} package in \texttt{R}:
$$\text{reduced} \sim \text{browser} + \text{outlet\_lean} + \text{topic} + \text{log(article\_length)} + (1 \mid \text{outlet}) + (1 \mid \text{article\_id})$$

For each measure of negative affect, the model is fit to summaries of negative articles, and \textit{reduced} is a binary outcome indicating whether the summary exhibits reduced negativity relative to its source article. Browser (reference: \textit{Chrome}), outlet leaning (reference: \textit{Center}), and topic (reference: \textit{Miscellaneous}) are included as categorical fixed effects. Standardized article length (log-scaled) is included as a control. Random intercepts for outlet and article account for non-independence among summaries for the same outlet and the same source article, with articles nested within outlets. The results are shown in Table~\ref{tab:reg-emo}. 

\begin{table}[t]
\centering
\small
\caption{Mixed-effects logistic regression of reduced negative affect, fit separately for each measure. Estimates on the logit scale. Reference categories: Chrome (browser), Center (outlet leaning), Miscellaneous (topic). Random intercepts: outlet, article. Negativity: $N=20{,}277$, anger: $N=24{,}594$, fear: $N=32{,}079$.
$^{*}p<.05$, $^{**}p<.01$, $^{***}p<.001$.}
\label{tab:reg-emo}
\begin{tabular}{l @{\hspace{20pt}} S[table-format=-1.3,table-column-width=1.1cm] @{}l S[table-format=1.3,table-column-width=1.1cm] S[table-format=-1.3,table-column-width=1.1cm] @{}l S[table-format=1.3,table-column-width=1.1cm] S[table-format=-1.3,table-column-width=1.1cm] @{}l S[table-format=1.3,table-column-width=1.1cm]}
\toprule
 & \multicolumn{3}{c}{Negativity} & \multicolumn{3}{c}{Anger} & \multicolumn{3}{c}{Fear} \\
\cmidrule(lr){2-4} \cmidrule(lr){5-7} \cmidrule(lr){8-10}
{Predictor} & \multicolumn{2}{c}{Estimate} & {SE} & \multicolumn{2}{c}{Estimate} & {SE} & \multicolumn{2}{c}{Estimate} & {SE} \\
\midrule
Intercept & -0.414 & & 0.313 & -0.065 & & 0.239 & -6.377 & $^{***}$ & 0.321 \\
\addlinespace
\multicolumn{10}{l}{\textit{Browser}} \\
\quad Comet & -0.509 & $^{***}$ & 0.051 & -0.957 & $^{***}$ & 0.046 & -0.566 & $^{***}$ & 0.061 \\
\quad Edge  & -0.121 & $^{*}$   & 0.049 &  1.712 & $^{***}$ & 0.051 &  1.157 & $^{***}$ & 0.061 \\
\addlinespace
\multicolumn{10}{l}{\textit{Outlet leaning}} \\
\quad Left  & -0.484 &         & 0.369 & -0.180 & & 0.260 & 0.229 &          & 0.129 \\
\quad Right & -0.748 & $^{*}$  & 0.371 &  0.049 & & 0.261 & 1.086 & $^{***}$ & 0.145 \\
\addlinespace
\multicolumn{10}{l}{\textit{Topic}} \\
\quad Abortion          & -0.330 &          & 0.348 &  0.521 &          & 0.315 &  0.616 &          & 0.534 \\
\quad Courts            &  0.214 &          & 0.289 &  0.903 & $^{***}$ & 0.266 &  1.649 & $^{**}$  & 0.580 \\
\quad Crime             &  1.399 & $^{***}$ & 0.237 &  0.673 & $^{**}$  & 0.216 & -0.471 &          & 0.326 \\
\quad Economy           &  0.006 &          & 0.219 &  0.685 & $^{**}$  & 0.208 & -0.360 &          & 0.301 \\
\quad Education         & -0.429 &          & 0.251 &  0.171 &          & 0.224 & -0.395 &          & 0.375 \\
\quad Election          & -0.124 &          & 0.203 &  0.597 & $^{**}$  & 0.184 &  1.050 & $^{**}$  & 0.322 \\
\quad Environment       & -0.442 &          & 0.276 &  0.430 &          & 0.260 & -0.483 &          & 0.375 \\
\quad Gender \& DEI     &  0.429 &          & 0.282 & -0.018 &          & 0.253 &  0.449 &          & 0.442 \\
\quad Government        & -0.261 &          & 0.206 &  0.312 &          & 0.186 &  0.414 &          & 0.309 \\
\quad Guns              &  0.784 & $^{*}$   & 0.356 &  0.228 &          & 0.335 & -0.105 &          & 0.535 \\
\quad Health            & -0.672 & $^{**}$  & 0.246 &  0.444 & $^{*}$   & 0.223 &  0.002 &          & 0.340 \\
\quad Immigration       &  0.177 &          & 0.246 &  0.435 &          & 0.222 & -0.556 &          & 0.348 \\
\quad Israel-Mideast    &  0.424 & $^{*}$   & 0.212 &  0.483 & $^{*}$   & 0.192 & -0.618 & $^{*}$   & 0.297 \\
\quad Religion          & -0.395 &          & 0.294 &  0.576 & $^{*}$   & 0.273 &  2.035 & $^{***}$ & 0.570 \\
\quad Russia-Ukraine    &  0.525 & $^{*}$   & 0.265 &  0.223 &          & 0.241 & -0.511 &          & 0.342 \\
\quad Technology        & -0.329 &          & 0.246 &  0.480 & $^{*}$   & 0.224 & -0.206 &          & 0.345 \\
\quad Trump Legal Cases &  0.473 &          & 0.264 &  0.471 & $^{*}$   & 0.240 &  2.734 & $^{***}$ & 0.579 \\
\quad World Affairs     & -0.117 &          & 0.211 &  0.314 &          & 0.192 & -0.478 &          & 0.293 \\
\addlinespace
\multicolumn{10}{l}{\textit{Control}} \\
\quad Article length & -0.069 & & 0.040 & 0.476 & $^{***}$ & 0.041 & 0.370 & $^{***}$ & 0.063 \\
\addlinespace
\midrule
\multicolumn{1}{l}{\textit{Random effects}} & \multicolumn{3}{c}{SD} & \multicolumn{3}{c}{SD} & \multicolumn{3}{c}{SD} \\
\quad Article & \multicolumn{3}{c}{2.254} & \multicolumn{3}{c}{2.431} & \multicolumn{3}{c}{6.043} \\
\quad Outlet  & \multicolumn{3}{c}{0.564} & \multicolumn{3}{c}{0.384} & \multicolumn{3}{c}{0.000} \\
\bottomrule
\end{tabular}
\end{table}

\paragraph{Qualitative analysis of summaries with increased negative affect}
To probe the reasons and implications of increased negative affect in AI news summaries, we conduct a qualitative analysis of 70 no-fear articles with increased fear in all three summaries, as well as 159 neutral-sentiment articles with increased negativity in all three summaries. The analysis first reveals that increased fear in summaries most commonly arises as criticism, tension, or threat becomes more directly conveyed through condensed claims and weakened hedging. For example, for an MSNBC article that discusses the ``DOGE dividend checks'' across 20 paragraphs with contextual buildup, quotes from both sides, and detailed math, summaries open with short, explicit claims such as \textit{The article explains that promised DOGE dividend checks are almost certainly not going to happen and were never realistic.} Similarly, a Hills article that illustrates a contested Democratic primary with rich and nuanced strokes receives summaries that explicitly characterize the reported event with fear-laden words such as \textit{flashpoint}, \textit{conflict}, or \textit{tension}. In an interesting exceptional case, fear increases in summaries where the order of events is reversed: whereas the article reads: \textit{A fire at ... was brought under control. The fire started at ...}, the summaries read: \textit{A fire broke out at  ... The fire was brought under control.} Summaries with increased negativity generally exhibit similar patterns to those with increased fear, in that themes of skepticism and criticism are surfaced and condensed through summarization, while the mitigating context is largely dropped.

Our findings suggest that increased negative affect in AI news summaries arises primarily from a skewed presentation of source facts. Although it is reassuring that the effect is not driven by added or fabricated information, the findings alert us to the risk that AI-based news summarizers may shape perception of the news in a negatively skewed way.

\paragraph{Change in negative affect across outlet leanings}
As seen in Fig.~\ref{fig:main-neg}B, differences across outlet leanings are most prominent for negativity and fear: negative articles from right-leaning outlets are least likely to receive less negative summaries (31\%/34\%/41\% for right/left/center), while fear-expressing articles from right-leaning outlets are most likely to receive summaries with less fear (27\%/21\%/17\% for right/left/center). Relative to centrist outlets, these differences are statistically significant for right-leaning outlets (p<.001 for fear, p<.05 for negativity), but not for left-leaning outlets. 

\paragraph{Change in negative affect across topics}
Fig.~\ref{fig:topic}B illustrates the change in negative affect across topics. The most exceptional pattern is that for threat-centered topics including Immigration, Israel-Mideast, and Crime, AI summarizers introduce fear for no-fear articles more often than reduce fear for fear-expressing articles (introduced/reduced: 22\%/17\% for Immigration, 20\%/18\% for Israel-Mideast, 18\%/17\% for Crime). While a two-proportion z-test shows that none of the differences are near significant, these topics stand in stark contrast with others including Religion, Trump Legal Cases, and Election, where fear reduction rates substantially exceed introduction rates (introduced/reduced: 6\%/36\% for Religion, 10\%/37\% for Trump Legal Cases, 8\%/30\% for Election). 
In addition, anger reduction rates are relatively comparable across topics (51--63\%), while negativity reduction rates are more variable (25--55\%). Relative to the Miscellaneous baseline, negative articles on Crime, Guns, Russia-Ukraine, and Israel-Mideast are significantly more likely to receive neutral summaries (p<.001 for Crime, p<.05 for the others). 

\subsection{Journalistic Writing Quality}\label{app:sec:res-qual}
\paragraph{Score transitions in journalistic writing quality}
Counts of article-to-summary score transitions for each browser and measure are visualized in Fig.~\ref{fig:transition}.

\paragraph{Statistical test of increase in journalistic writing quality}
We test for the effect of browser, article leaning, and topic on increased journalistic writing quality by AI summarizers. For each measure of journalistic writing quality (i.e., \textit{clarity}, \textit{engagement}, \textit{personal tone} and \textit{clickbait}), we fit the following mixed-effects logistic regression model using the \texttt{glmmTMB} package in \texttt{R}:
$$\text{increased} \sim \text{browser} + \text{outlet\_lean} + \text{topic} + \text{log(article\_length)} + (1 \mid \text{outlet}) + (1 \mid \text{article\_id})$$

For each measure of journalistic writing quality, the model is fit to summaries of low-quality articles---respectively, articles labeled \textit{clarity} 0, \textit{engagement} 0 or 1, \textit{personal tone} 1 or 2, and \textit{clickbait} 1 or 2. \textit{Increased} is a binary outcome indicating whether the summary exhibits increased journalistic writing quality---respectively, increased clarity, increased engagement, decreased personal tone, or decreased clickbait tendency---relative to its source article. Browser (reference: \textit{Chrome}), outlet leaning (reference: \textit{Center}), and topic (reference: \textit{Miscellaneous}) are included as categorical fixed effects. Standardized article length (log-scaled) is included as a control. Random intercepts for outlet and article account for non-independence among summaries for the same outlet and the same source article, with articles nested within outlets. The results are shown in Table~\ref{tab:reg-qual}. 

\begin{table}[t]
\centering
\small
\caption{Mixed-effects logistic regression of increased journalistic writing quality, fit separately for each measure. Estimates on the logit scale. Reference categories: Chrome (browser), Center (outlet leaning), Miscellaneous (topic). Random intercepts: outlet, article. Clarity: $N=3{,}852$, engagement: $N=34{,}416$, personal tone: $N=9{,}105$, clickbait: $N=804$.
$^{*}p<.05$, $^{**}p<.01$, $^{***}p<.001$. Coefficients marked with † arise from complete or quasi-complete separation and are not interpretable. Empty cells indicate no qualifying articles on the corresponding topics.}
\label{tab:reg-qual}
\begin{tabular}{l @{\hspace{14pt}} S[table-format=-2.3,table-column-width=1.2cm] @{}l S[table-format=1.3,table-column-width=1.0cm] S[table-format=-2.3,table-column-width=1.2cm] @{}l S[table-format=1.3,table-column-width=1.0cm] S[table-format=-2.3,table-column-width=1.2cm] @{}l S[table-format=1.3,table-column-width=1.0cm] S[table-format=-2.3,table-column-width=1.2cm] @{}l S[table-format=1.3,table-column-width=1.0cm]}
\toprule
 & \multicolumn{3}{c}{Clarity} & \multicolumn{3}{c}{Engagement} & \multicolumn{3}{c}{Personal} & \multicolumn{3}{c}{Clickbait} \\
\cmidrule(lr){2-4} \cmidrule(lr){5-7} \cmidrule(lr){8-10} \cmidrule(lr){11-13}
{Predictor} & \multicolumn{2}{c}{Estimate} & {SE} & \multicolumn{2}{c}{Estimate} & {SE} & \multicolumn{2}{c}{Estimate} & {SE} & \multicolumn{2}{c}{Estimate} & {SE} \\
\midrule
Intercept & 3.447 & $^{***}$ & 0.597 & -11.054 & $^{***}$ & 0.578 & 2.014 & $^{***}$ & 0.276 & 2.493 & & 1.609 \\
\addlinespace
\multicolumn{13}{l}{\textit{Browser}} \\
\quad Comet & 0.303 & $^{*}$   & 0.145 & 1.837 & $^{***}$ & 0.169 & 0.206 & $^{*}$   & 0.087 & -0.881 & & 0.458 \\
\quad Edge  & -0.581 & $^{***}$ & 0.138 & -0.472 & $^{**}$ & 0.167 & -1.044 & $^{***}$ & 0.086 & -0.881 & & 0.458 \\
\addlinespace
\multicolumn{13}{l}{\textit{Outlet leaning}} \\
\quad Left  & 0.362 & & 0.512 & -0.220 & & 0.339 & -0.539 & $^{**}$ & 0.201 & 2.501 & $^{*}$ & 1.090 \\
\quad Right & 0.813 & & 0.498 & -0.125 & & 0.336 & -0.453 & $^{*}$  & 0.190 & 2.806 & $^{*}$ & 1.106 \\
\addlinespace
\multicolumn{13}{l}{\textit{Topic}} \\
\quad Abortion          & -0.678 &          & 1.351 & -1.254 &          & 1.782 &  0.981 &          & 0.698 &  0.562 &                 & 2.915 \\
\quad Courts            & -1.024 &          & 0.752 & -1.245 &          & 1.353 &  1.061 &          & 0.588 & 30.750 & $^{\dagger}$ & \multicolumn{1}{c}{--} \\
\quad Crime             & -1.126 &          & 0.757 &  0.768 &          & 0.639 &  0.663 & $^{*}$   & 0.324 &        &                 & \\
\quad Economy           & -3.119 & $^{***}$ & 0.516 & -0.357 &          & 0.646 &  2.351 & $^{***}$ & 0.506 &  2.874 &                 & 1.694 \\
\quad Education         & -0.508 &          & 0.856 & -0.469 &          & 0.922 &  1.717 & $^{***}$ & 0.429 &  1.160 &                 & 2.627 \\
\quad Election          & -0.231 &          & 0.524 & -0.489 &          & 0.653 &  1.381 & $^{***}$ & 0.307 &  4.067 & $^{*}$          & 1.781 \\
\quad Environment       & -1.175 &          & 0.657 & -0.902 &          & 1.091 &  2.022 & $^{***}$ & 0.565 &  2.842 &                 & 2.438 \\
\quad Gender \& DEI     &  2.155 &          & 1.409 & -0.385 &          & 1.057 &  0.391 &          & 0.415 &  3.441 &                 & 2.244 \\
\quad Government        & -1.052 &          & 0.543 & -0.663 &          & 0.695 &  1.655 & $^{***}$ & 0.405 &  3.766 & $^{*}$          & 1.750 \\
\quad Guns              & 16.299 & $^{\dagger}$ & \multicolumn{1}{c}{--} & 0.365 &          & 1.037 &  0.335 &          & 0.543 &        &                 & \\
\quad Health            & -0.636 &          & 0.717 & -0.932 &          & 0.920 &  1.700 & $^{***}$ & 0.440 &  2.296 &                 & 1.917 \\
\quad Immigration       &  0.728 &          & 0.948 & -0.444 &          & 0.871 &  1.709 & $^{***}$ & 0.429 & 28.730 & $^{\dagger}$ & \multicolumn{1}{c}{--} \\
\quad Israel-Mideast    & -0.079 &          & 0.604 & -0.227 &          & 0.687 &  0.535 &          & 0.343 &  1.942 &                 & 2.547 \\
\quad Religion          & -0.108 &          & 0.600 &  0.410 &          & 0.904 & -1.019 & $^{**}$  & 0.347 &  0.061 &                 & 1.889 \\
\quad Russia-Ukraine    &  2.220 &          & 1.313 &  0.287 &          & 0.762 &  1.391 & $^{**}$  & 0.530 &  1.877 &                 & 1.903 \\
\quad Technology        &  0.259 &          & 0.656 & -1.198 &          & 1.015 &  1.406 & $^{**}$  & 0.430 &  2.923 &                 & 1.938 \\
\quad Trump Legal Cases &  0.237 &          & 1.083 & -1.000 &          & 1.099 &  1.444 & $^{**}$  & 0.517 & 27.570 & $^{\dagger}$ & \multicolumn{1}{c}{--} \\
\quad World Affairs     & -0.466 &          & 0.567 & -0.655 &          & 0.700 &  1.344 & $^{***}$ & 0.347 &  3.126 &                 & 2.369 \\
\addlinespace
\multicolumn{13}{l}{\textit{Control}} \\
\quad Article length & 0.029 & & 0.108 & 0.174 & & 0.116 & 0.637 & $^{***}$ & 0.076 & -0.111 & & 0.448 \\
\addlinespace
\midrule
\multicolumn{1}{l}{\textit{Random effects}} & \multicolumn{3}{c}{SD} & \multicolumn{3}{c}{SD} & \multicolumn{3}{c}{SD} & \multicolumn{3}{c}{SD} \\
\quad Article & \multicolumn{3}{c}{2.131} & \multicolumn{3}{S[table-format=2.3]}{11.130} & \multicolumn{3}{c}{3.027} & \multicolumn{3}{c}{5.490} \\
\quad Outlet  & \multicolumn{3}{c}{0.661} & \multicolumn{3}{S[table-format=2.3]}{0.000}  & \multicolumn{3}{c}{0.000} & \multicolumn{3}{c}{0.000} \\
\bottomrule
\end{tabular}
\end{table}

\begin{figure}[h]
\centering
\includegraphics[width=\linewidth]{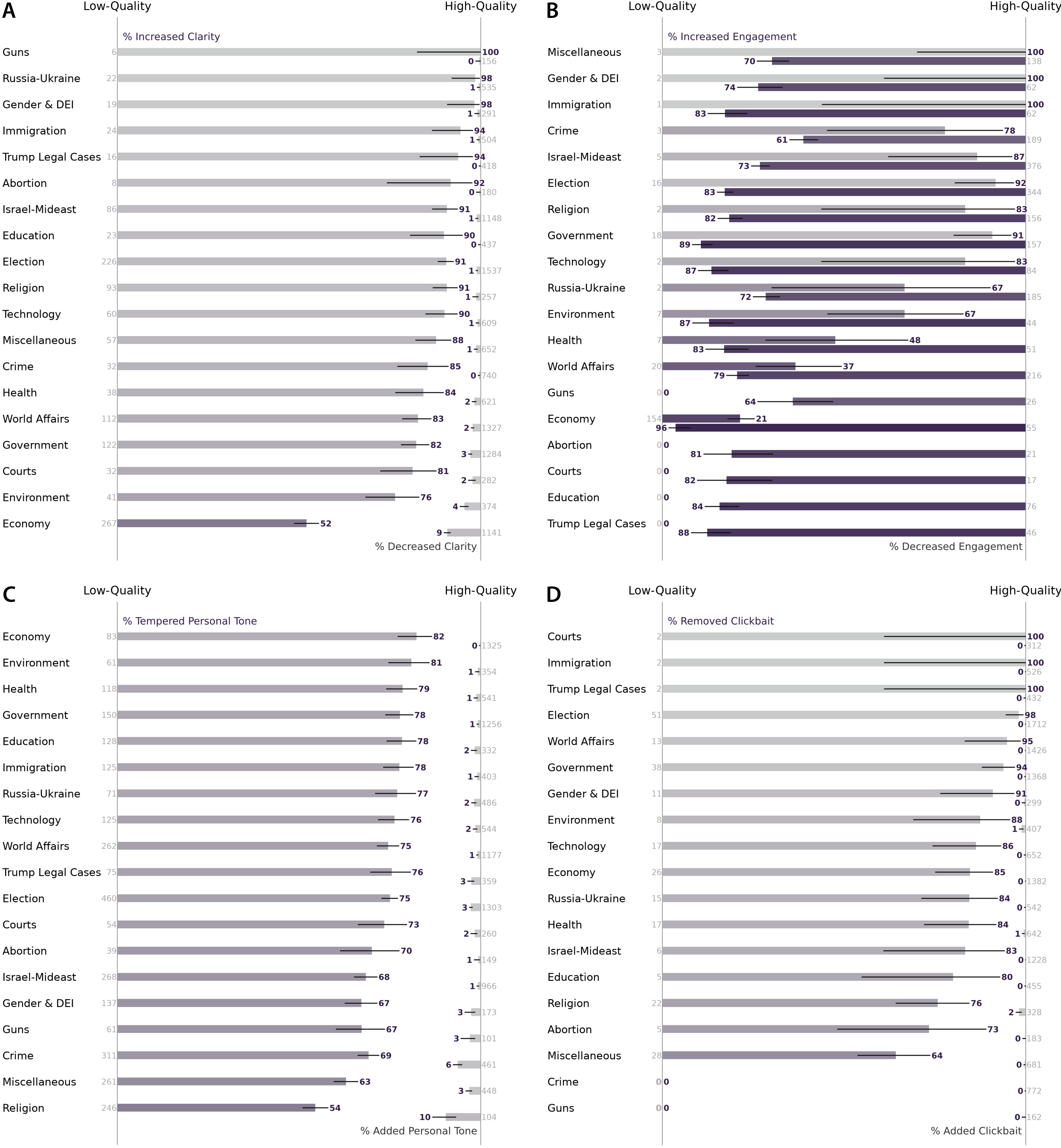}
\caption{Change in journalistic writing quality by topic, as measured by (A) \textit{clarity}, (B) \textit{engagement}, (C) \textit{personal tone}, and (D) \textit{clickbait}. In each panel, topics are ordered by the \textit{quality shift} they induce, from most quality-increasing (top) to lowest (bottom). The quality shift is calculated as the share of low-quality articles receiving higher-quality summaries minus the share of high-quality articles receiving lower-quality summaries. In each panel, bars extending from the left show the share of low-clarity/low-engagement/high-personal/high-clickbait articles receiving summaries of higher quality, and bars extending from the right show the share of high-clarity/high-engagement/low-personal/low-clickbait articles receiving summaries of lower quality. Small numbers within or besides bars are article counts. Error bars indicate 95\% confidence intervals.}
\label{fig:qual-topic}
\end{figure}

\paragraph{Change in journalistic writing quality across topics}
Fig.~\ref{fig:qual-topic} illustrates the change in journalistic writing quality across topics. Most prominently, low-clarity articles on Economy are far less likely than the Miscellaneous baseline to receive summaries with increased clarity (52\% vs. 88\%, p<.001). In addition, personal tone is tempered slightly less often for articles from partisan outlets than those from centrist ones (72\%/70\%/73\% for left/right/center, p<.01 for left-leaning, p<.05 for right-leaning). Across topics, personal tone is tempered more often for articles on Economy (82\%, p<.001), Environment (81\%, p<.001), and Education (78\%, p<.001), and less often for articles on Religion (54\%, p<.01), compared to the Miscellaneous baseline (63\%).

\begin{figure}[h]
\centering
\includegraphics[width=\linewidth]{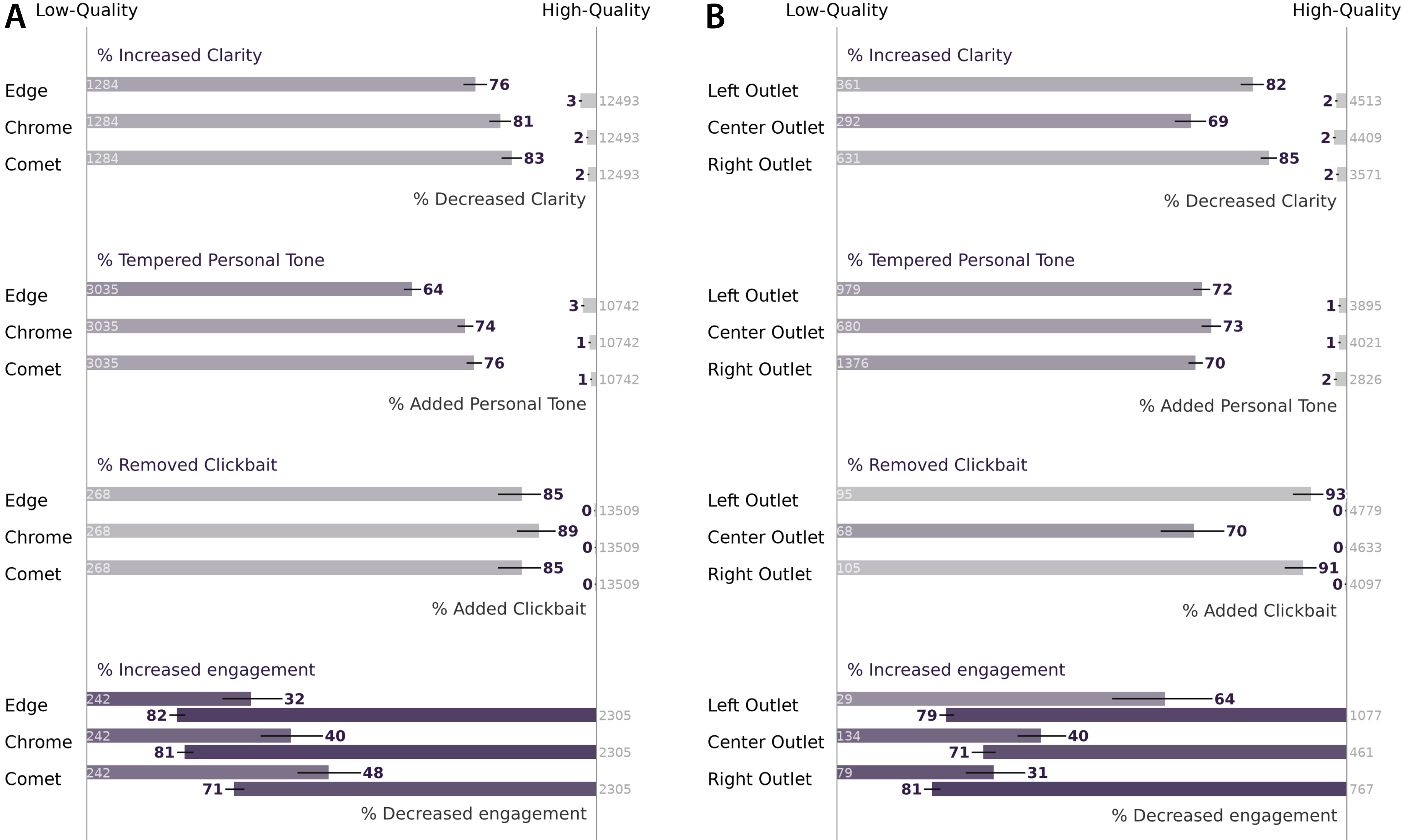}
\caption{Change in \textit{clarity}, \textit{personal tone}, \textit{clickbait}, and \textit{engagement} introduced by AI news summarizers, by (A) browser and (B) outlet leaning. In each panel, bars extending from the left show the share of low-clarity/high-personal/high-clickbait/low-engagement articles receiving summaries of higher quality, and bars extending from the right show the share of high-clarity/low-personal/low-clickbait/high-engagement articles receiving summaries of lower quality. Small numbers within or besides bars are article counts. Error bars indicate 95\% confidence intervals.}
\label{fig:main-quality}
\end{figure}

\end{document}